\documentstyle[preprint,eqsecnum,aps]{revtex}
\input epsf

\tightenlines
 
\def\INSERTCAP#1#2{\vbox{\multiply\baselineskip
by 3 \divide\baselineskip by 4
{
\noindent%
{\rm Table #1. }{\sl #2 \medskip}}
}\multiply\baselineskip by 4 \divide\baselineskip by 3 \hfill}

\def\INSERTFIG#1#2#3{\multiply\baselineskip
by 3 \divide\baselineskip by 4
\vbox{\bigskip \vbox{\hfil\epsfbox{#1}\hfill}%
{
\noindent%
{FIG.\ #2. }{\sl #3 \medskip}}
}\multiply\baselineskip by 4 \divide\baselineskip by 3 \hfill}%
\def\INSERTtwoFIGs#1#2#3#4{\multiply\baselineskip
by 3 \divide\baselineskip by 4
\vbox{\bigskip \vbox{\hfil\epsfbox{#1}\hfill\epsfbox{#2}\hfill}%
{
\noindent%
{FIG.\ #3. }{\sl #4 \medskip}}
}\multiply\baselineskip by 4 \divide\baselineskip by 3 \hfill}%
\begin{document}

\draft
\preprint{\vbox{
\hbox{UCSD/PTH 98--14} \hbox{JLAB-THY-98--18}
}}

\title{Quark-Hadron Duality in the 't~Hooft Model for Meson Weak
Decays: Different Quark Diagram Topologies}

\author{Benjam\'{\i}n Grinstein\footnote{bgrinstein@ucsd.edu}}

\address{Department of Physics,
University of California at San Diego, La Jolla, CA 92093}

\author{Richard F. Lebed\footnote{lebed@jlab.org}}

\address{Jefferson Lab, 12000 Jefferson Avenue, Newport News, VA
23606}

\date{May, 1998}

\maketitle
\begin{abstract}
\tightenlines
	We compare the effects of different quark diagram topologies
on the weak hadronic width of heavy-light mesons 
in the large $N_c$ limit.  
We enumerate the various topologies and show that the
only one dominant (or even comparable) in powers of $N_c$ to the
noninteracting spectator ``tree'' diagram is the ``annihilation''
diagram, in which the valence quark-antiquark pair annihilate weakly.
We compute the amplitude for this diagram in the
't~Hooft model (QCD in 1+1 spacetime dimensions with a large number of
colors $N_c$) at the hadronic level
and compare to the Born term partonic level. We find that 
quark-hadron duality is not
well satisfied, even after the application of a smearing procedure to
the hadronic result.  A number of interesting subtleties absent from
the tree diagram case arise in the annihilation diagram case, and are
described in detail.  
\end{abstract}



\narrowtext

\section{Description of the Problem} \label{intro}

	The notorious difficulty of computing strong interaction
quantities from first principles leads directly to a multitude of
models, methods, and approximations.  QCD lore dictates that inclusive
rates, {\it i.e.}, those for which we do not inquire about any detail
of the final hadronic state, are given to good approximation by the
corresponding rate in a theoretical world of unconfined, perturbative
final-state quarks.  This lore, known as ``quark-hadron duality'' is
seldom shown to follow from first principles.  In some cases, such as
the rate of $e^+e^-\to$ hadrons, an operator product expansion
can be used to demonstrate the lore for quantities in which one
averages over the energy of the final-state hadronic system.  This is
the so-called ``global'' duality, in contrast to duality for
unaveraged quantities, or ``local'' duality.

	Quark-hadron duality applied to decays of heavy mesons permits
the computation of many important quantities, such as meson lifetimes,
hadronic branching fractions, and the average number of charm quarks
per beauty quark decay.  One must note that the duality used here is
of the ``local'' variety, and therefore less likely to be valid.
Indeed, the sizeable deviations between theoretical predictions utilizing this
form of duality and experiment for lifetimes, branching fractions and
charm countings suggest that either there are large corrections to
duality, or worse, that the duality lore simply does not apply in this
context.

	In this paper we study whether duality holds for annihilation
decays of heavy mesons in a highly simplified but soluble strongly
interacting theory; in a previous paper\cite{GL} we explored this
issue for spectator decays of heavy mesons.  We also investigate the
validity of duality in the interference between spectator and
annihilation decays.

	To inquire about duality in a model, one must require that the
model exhibits permanent quark confinement and asymptotic freedom; the
former is necessary for a meaningful definition of purely hadronic
properties, while the latter is believed to be the main ingredient of
duality.  The 't~Hooft model is a good candidate: It satisfies these
requirements and behaves, in many other respects, much like the real
world of strong interactions.

	Perhaps still the most straightforward and frequently used
model of strong interactions is the constituent quark model, in which
hadrons are envisioned as consisting of confined but otherwise weakly
interacting valence quarks.  Although deep inelastic scattering
experiments clearly show that the true structure of a hadron
incorporates a much more complicated brew of gluons and sea quarks,
the undeniable success of the quark model in predicting hadronic
spectra or enumerating decay modes still leads researchers to apply
this scheme immediately when approaching a new hadronic system.

	The asymptotically free parton result, {\it i.e.}, the result
of the valence quark model dressed with perturbative QCD corrections,
is also the leading term in an Operator Product Expansion\cite{Wils}
(OPE), in those cases where an OPE is known to exist, such as deep
inelastic lepton-hadron scattering\cite{DIS} and $e^+ e^-$
annihilation\cite{SVZ}, as well as semileptonic processes involving
heavy quarks\cite{CGG}.  One feature held in common by these processes
is the presence of a large mass or energy scale, which provides the
inverse expansion parameter of the OPE; consequently, it is tempting
to suppose that other processes with large scales also possess OPEs.
Such is the case for the nonleptonic decays of $B$ mesons, where an
OPE in powers of $1/m_b$ is purported to exist\cite{bigi}.

 	As of the time of this writing, measurements of many of these
exclusive channels at CLEO and LEP are being performed for the first
time.  Apart from the intrinsic value of such information, nonleptonic
$B$ decays are expected to provide valuable insights into QCD,
magnitudes of CKM elements, and CP violation; consequently,
understanding these decays has become the focus of much recent
theoretical work.  But the theoretical situation is wide open,
precisely because no part of the decay is free of the complications of
strong interaction physics.

	When applied to a $\bar B$ meson, the lore of quark model
calculations or the expansion of Ref.~\cite{bigi} declares that the
inclusive decay width should be dominated by the (color-unsuppressed)
``tree'' diagram T (Fig.~1), in which the $b$ quark decays to a
lighter flavor by emission of a $W^-$, particularly as $m_b \to
\infty$.  The daughter quark of the $b$ then combines with the
spectator antiquark to form one meson, whereas the quark-antiquark
pair from the nonleptonic decay of the $W$ form another.  The above
process assumes factorization, which means that the $W$ system and
daughter-spectator systems are regarded as non-interacting after the
initial weak vertex.  One very interesting comparison that can be made
at this point is between the width obtained by the sum of such
diagrams in Fig.~1 regarded as hadronic decays and the corresponding
free quark decay.
\INSERTFIG{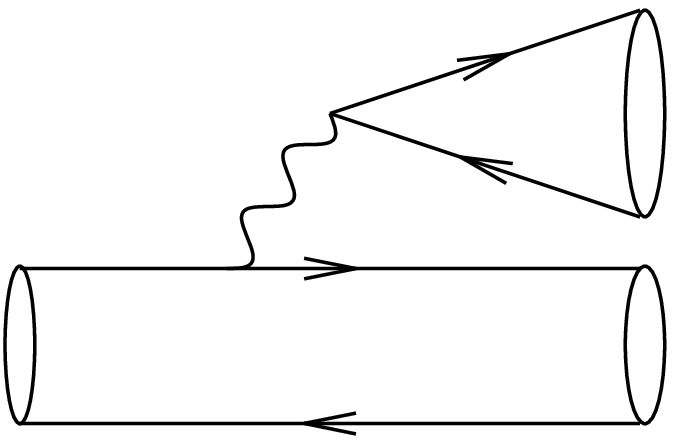}{1}{The (color-unsuppressed) ``tree'' (T) parton
diagram for the decay of one meson into two mesons.  Ovals indicate
the binding of partons into hadrons.}

	Of course, such a diagram is but one possibility, even in the
simple quark model.  For example, the spectator antiquark and the
antiquark from the $W^-$ decay may trade places before hadronization,
the ``color-suppressed'' tree diagram C (Fig.~2).  Such an amplitude 
is suppressed by a factor of
$N_c$, the number of QCD color charges, compared to the T amplitude,
since the $W$ is a color singlet; thus, the color indices are
automatically suitable for creating colorless mesons in Fig.~1, but
require rearrangement in Fig.~2.  Of course, in a real meson there may
be additional dynamical enhancements or suppressions, and moreover, 
the exchange of {\em any}
number of gluons and color charge between the quarks on opposite sides
of the $W$ line may completely muddle the hierarchy based on large $N_c$
counting.
\INSERTFIG{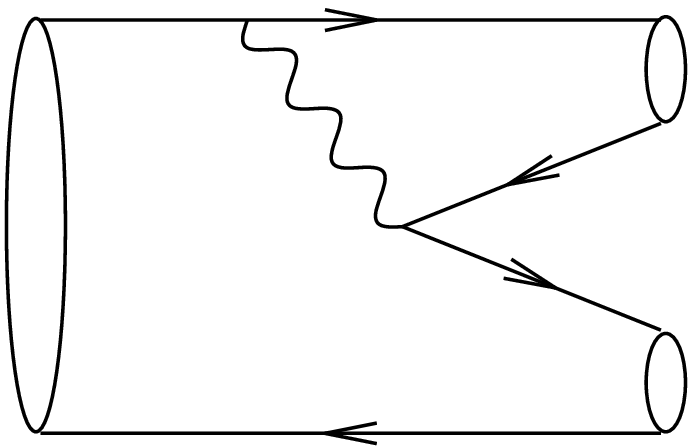}{2}{The ``color-suppressed'' (C) parton diagram
for the decay of one meson into two mesons.  Ovals indicate the
binding of partons into hadrons.}

	Likewise, the ``annihilation'' amplitude A (Fig.~3), in which
the valence quark-antiquark pair of the $\bar B$ annihilate to a $W$
(quantum numbers permitting), is assumed to be suppressed compared to
the T amplitude because of the difficulty of the $b$ quark and the
antiquark ``finding'' each other in the meson in order to annihilate.
Quantitatively, this probability is proportional to the square
of the meson wave function at vanishing quark separation, $|\psi(0)|^2
\propto f_B^2$ (the van Royen-Weisskopf relation).  To compare this
probability with that of the T
diagram, one then argues that the only remaining dimensionful quantity
that can be used to form a probability is $m_B$, so that the
relative probability of an annihilation to a tree process is
$f_B^2/m_B^2 \leq O(0.2\%)$.
\INSERTFIG{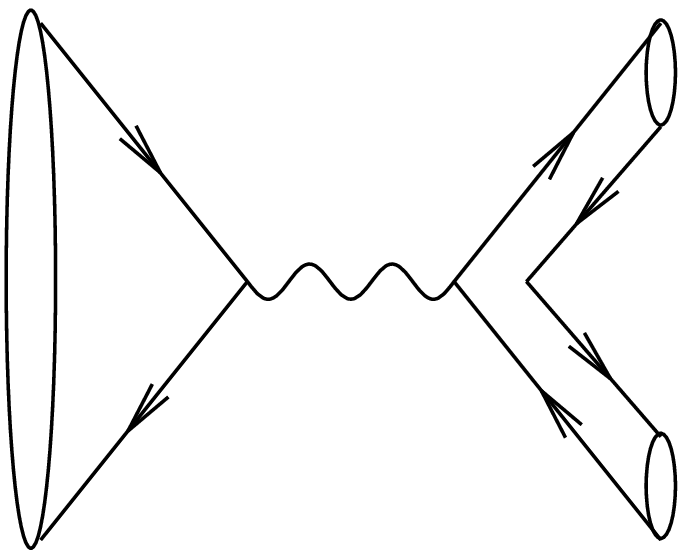}{3}{The ``annihilation'' (A) parton diagram for
the decay of one meson into two mesons.  Ovals indicate the binding of
partons into hadrons.}

	Nevertheless, other concerns lead one to believe that the
annihilation and other non-spectator diagrams may be more important
than one would naively expect.  For, if non-interacting spectator
diagrams dominate decays of $b$ hadrons, how then does one explain the
fact that the lifetime of $\bar B$ mesons is greater than that of
$\Lambda_b$ baryons by 30\% or more\cite{tLam}?  One possible
explanation is that the naive estimate of the previous paragraph fails
to include potentially large numerical coefficients.  This point of
view has been voiced recently in Ref.~\cite{NS}, in which it is
suggested that the annihilation width has an unexpected additional
enhancement of 16$\pi^2$ compared to the tree width.  One obtains this
result by application of perturbative unitarity of the S-matrix to a
cut across loop diagrams (Figs.\ 4$a$,$b$).
\INSERTFIG{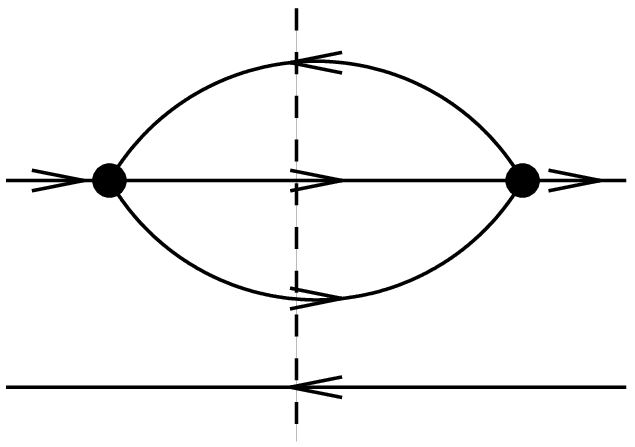}{4$a$}{Diagram giving rise to the ``tree''
amplitude of Fig.~1 upon a vertical cut through the center
(application of unitarity).  The vertex blobs indicate $W$ exchange.}
\INSERTFIG{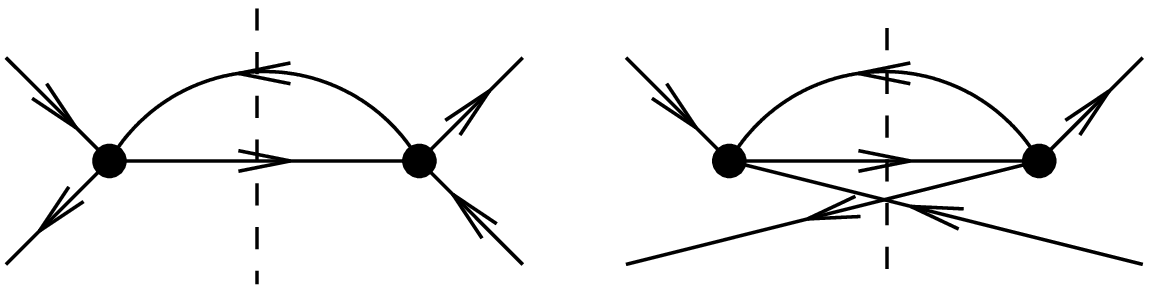}{4$b$}{Diagrams giving rise to the
``annihilation'' amplitude of Fig.~3 upon a vertical cut through the
center (application of unitarity).  The vertex blobs indicate $W$
exchange.  Strongly produced $q\bar q$ pairs are not drawn here for
simplicity.}
Since the diagram giving rise to the tree width requires an additional
loop integration compared to that for the annihilation width, the
latter is enhanced by a relative factor of 16$\pi^2$ (multiplied by
the original factor $f_B^2/m_B^2$).

	But is such an effect genuine?  As soon as one includes the
gluon lines necessary to bind real mesons, the counting of loops no
longer leads to an obvious enhancement, unless we impose another
hierarchy to suppress the
effects of these exchanged gluons.  The difficulty originates, as
always, in our inability to take into account the multitude of gluons
and virtual quarks involved in the strong coupling of mesons.

	A question similar in spirit is that of the sense in which the
partonic annihilation diagram Fig.~3 exhibits quark-hadron duality
with its corresponding sum of exclusive hadronic channels.  Just as it
is expected that the width obtained from the tree diagram Fig.~1
approaches its corresponding sum of exclusives as the heavy quark mass
becomes large, one may study whether this is true for processes in
which the quantum numbers of the quarks are such that an annihilation
diagram but not a tree diagram is permitted.

	Unfortunately, the determination of definitive solutions to
these questions requires one to have in hand exact solutions of
QCD, and so is currently out of reach.

	However, there does exist a simpler universe in which QCD can
be solved exactly, so that we may consider the problem of meson decays
in either quark or hadron language.  As 't~Hooft showed long
ago\cite{tH}, the Green functions of QCD in one spatial and one time
dimension (1+1) in the limit of $N_c \to \infty$ are completely
calculable.  Despite these large modifications from our universe, the
't~Hooft model remains a nontrivial theory that possesses the
attractive feature of realizing confinement by binding quark-antiquark
pairs into color-singlet mesons (and more generally, forbidding
colored states\cite{CCG}), as well as asymptotic freedom and
the many phenomenological
results of large-$N_c$ QCD\cite{tHN} common to our $N_c = 3$ universe,
such as dominance of scattering amplitudes by diagrams with a minimum
of meson states, OZI suppression, the absence of exotics, and
others\cite{Wit}.

	In this paper we consider the analogue of $\bar B$ weak
nonleptonic decays in 1+1, where ``$\bar B$'' means a meson with a
heavy quark (``$b$'') of mass $M$ and a light antiquark of mass $m$;
``heavy'' and ``light'' quark are terms made more precise in
Sec.~\ref{review}.  We studied the question of quark-hadron duality
for the tree diagram in Ref.~\cite{GL}, in which it was shown that
agreement between the two pictures as $M \to \infty$ occurred in a
subtle and surprising manner with such high precision that the
discrepancy between the two yielded a remarkable result: a correction
to the Born term partonic limit well-fit numerically by a term
of relative order $1/M$.  Encouraged by this
result, we ask what may be learned from other topologies of Feynman
diagrams.  We study in detail the annihilation diagram and
compare the total width with that obtained from the parton model.
Our motivation, as before, is the hope that the results
from a soluble theory quite similar to real QCD in some respects and
quite different in others may shed light on the full four-dimensional
problem.

	The paper is organized as follows.  We begin in
Sec.~\ref{diags} with an enumeration of the various quark diagram
topologies and study the scaling of each in the large $N_c$ limit.
This classification is independent of the number of spacetime
dimensions.  We find that, in the decays of interest, naive $N_c$
power counting at the diagram level is subtle and deserves special
discussion.  The annihilation diagram emerges as the dominant topology
and is the one whose computation is studied in the subsequent portion
of the paper.  In Sec.~\ref{review}, we briefly review the more arcane
features of 1+1 dimensional physics and the 't~Hooft model that are
particularly relevant to the subsequent calculations.
Section~\ref{inc} exhibits the algebraic results of the inclusive
parton-level calculation of widths, from both a naive tree-level
diagram and an analysis based on loop diagrams as in Fig.~4$b$, the
latter being related closely to the corresponding OPE-like expansion.
In Sec.~\ref{exc}, we present the results of the width calculation
through exclusive hadronic channels in the 't~Hooft model.
Section~\ref{res} gives our numerical results and a discussion of
their implications, and Sec.~\ref{conc} concludes. A short Appendix
discusses the van Royen-Weisskopf relation in arbitrary dimensions.

\section{Quark Diagram Topologies and Large $N_c$} \label{diags}

	In order to obtain Green functions for exclusive decays in the
't~Hooft model, one must first decide which diagrams are present at
leading order in $N_c$.  As is well known, $n$-meson couplings in
large $N_c$ appear with a suppression factor $N_c^{(1-n/2)}$, and
therefore the leading meson decay diagrams are those producing only
two final-state mesons.  However, for some diagram topologies there is
the possibility of direct oscillation of the $\bar B$ meson into a
single highly excited meson of the same mass.  Such resonant
production poses an interesting problem of large $N_c$ counting, as we
discuss below.

	Next, we consider the relevant diagrams in terms of quarks,
gluons, and electroweak bosons, in order to count factors of $N_c$
appearing in these diagrams.  To lowest order in electroweak coupling,
a single gauge boson is required; moreover, since we require the decay
of a quark, the boson must be a flavor-changing $W$ rather than a
$\gamma$ or $Z$.  It is convenient to classify all possible diagram
topologies according to six classes given by Ref.~\cite{GHLR}, since
these account for all possible diagrams including only one electroweak
gauge boson.  The categories include the color-unsuppressed tree
diagram T (Fig.~1), the color-suppressed tree diagram C (Fig.~2), the
annihilation diagram A (Fig.~3), the electroweak ``exchange'' diagram
E (Fig.~5$a$), the ``penguin'' diagram P (Fig.~5$b$), and the
``penguin annihilation'' diagram PA (Fig.~5$c$).  For the P diagram
exhibited in Fig.~5$b$ it should be noted that the gluon may instead
attach directly to the spectator antiquark, in which case production
of an additional quark-antiquark pair is not required.  Similarly, the
PA topology includes diagrams in which the intermediate glue connects
to a single final-state $q\bar q$ pair.  Although these diagrams were
originally exhibited in the context of quark model calculations, they
suit our purposes since the inclusion of each additional meson,
quark-antiquark pair, or external gluon is accompanied by suppressions
in the amplitude of orders $1/\sqrt{N_c}$, $1/N_c$ and $1/\sqrt{N_c}$,
respectively.  Furthermore, diagrams with internal gluons are either
nonplanar and hence suppressed by powers of 1/$N_c^2$, or planar and
produce a diagram at most of the same order in $N_c$ as the diagram
obtained when all such gluons are removed.  Thus, the simple diagrams
displayed in the figures are representatives of those with the leading
behavior in $N_c$ for each possible topology.

	The electroweak gauge boson does not carry color charge, and
so the possible diagrams fall into two categories: those in which the
boson connects two otherwise disjoint color loops, and those in which
the boson begins and ends on quarks already connected by gluons and
therefore within a single color loop structure.  Clearly, the former
diagrams boast one extra color loop, and thus dominate the latter by a
factor of $N_c$.  The former set consists only of the T and A
diagrams, whereas the latter set includes C, E, P, and PA diagrams.
Only for T and A diagrams does the amplitude factorize into a product
of a decay constant and the matrix element of a current between two
mesons.  Finally, forming a color-singlet meson for the one initial
and two final mesons brings in a factor of $N_c^{-3/2}$ for each
amplitude, and so we find that the amplitude is $\propto N_c^{+1/2}$ for T
and A, $\propto N_c^{-1/2}$ for C, E, and P, and $\propto N_c^{-3/2}$
for PA.  Widths are obtained by squaring the amplitudes and folding in
phase space as usual, which adds no powers of $N_c$ since meson masses
and momenta scale as $N_c^0$.  For reasons that will presently become
transparent, let us refer to the above as ``naive'' $N_c$ power
counting.

\INSERTFIG{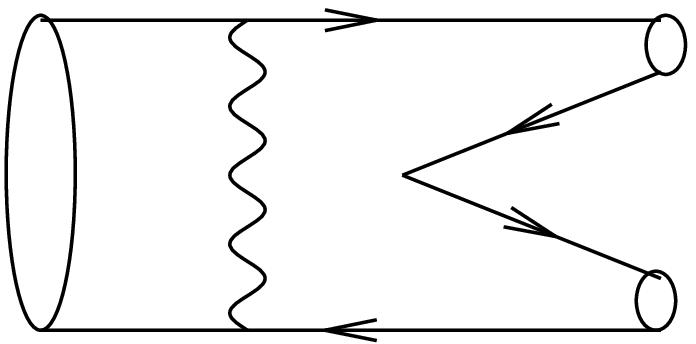}{5$a$}{Electroweak ``exchange'' (E) parton
diagram.}

\INSERTFIG{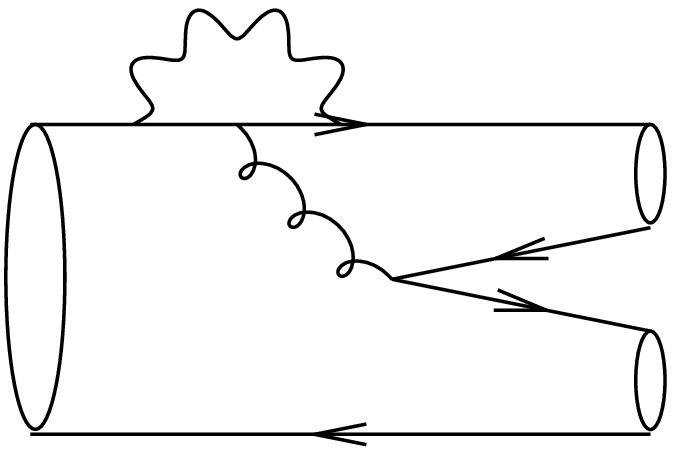}{5$b$}{``Penguin'' (P) parton diagram.}

\INSERTFIG{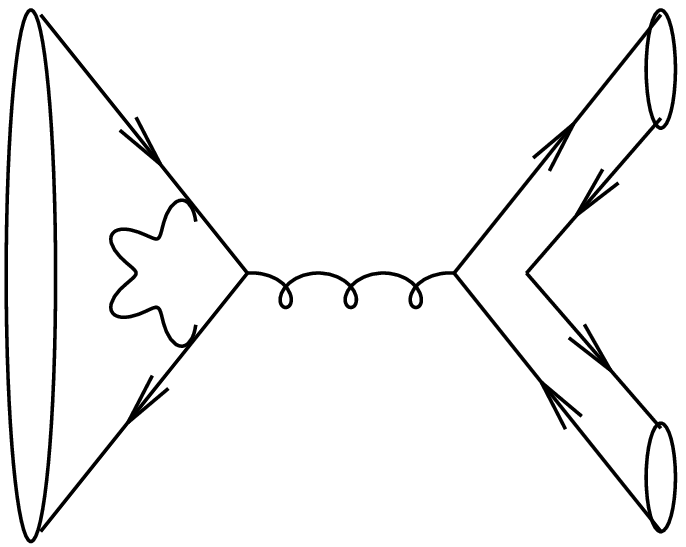}{5$c$}{``Penguin annihilation'' (PA) parton
diagram.  Since the initial and final states are color singlets, the
intermediate state actually requires at least two gluons; however, the
archetype presented here exhibits the same $N_c$ counting.}

	Suppose, however, that a given diagram topology permits a
resonant flavor-changing transition into a {\em single\/} meson, to
which we may assign the label ``one-meson decay.''  Such diagrams
clearly dominate over those of the corresponding two-meson decay by a
factor of $\sqrt{N_c}$ in amplitude.  Therefore, 't~Hooft model (or
indeed any large $N_c$) studies of the two-meson decay mode for such
diagrams appear doomed since the 't~Hooft model presents results only
for behavior at leading order in $N_c$.

	Nevertheless, the one-meson decay is a very strange physical
process.  For such a transition to occur on-shell, the final-state
light-quark meson must have precisely the same mass as the initial
heavy-light meson.  The weak decay width as a function of $M$ to
lowest order in $N_c$ is then a series of delta function spikes (since
the strong decay widths of the light quark mesons scale as $1/N_c$),
with a nonzero value if and only if $M$ is tuned just right to produce
such a light quark meson.  In this picture, a continuum width appears
only at relative order $(1/\sqrt{N_c})^2 = 1/N_c$, when two-body
decays are permitted.

	Clearly, this is an unsatisfactory physical picture.  It
implies that the leading behavior of the weak decay width scales
as $N_c^n$ for some integer $n$ if $M$ is tuned to certain special
values, but scales as $N_c^{n-1}$ otherwise.  The simple large $N_c$
diagram counting appears to have failed us.

	Fortunately, it is not difficult to develop a useful and
consistent physical interpretation for such situations.  The salient
point is to consider $N_c$ very large but still finite, so that strong
widths of light-quark mesons are not strictly zero; indeed, the usual
$N_c$ counting shows that their $O(1/N_c)$ strong widths are exactly
what one obtains from strong decays into two lighter mesons.  Thus,
the one-meson decays may be interpreted as intermediate states in
the decay of the initial heavy-light meson through a single resonance
into the final two-meson state.

	Effectively, in this picture one integrates out the
one-particle decay resonant channels, which dominate the nonresonant
two-particle weak decay widths by $N_c^1$, from the hadronic
Lagrangian.  However, unitarity demands that the extra power of $N_c$
must appear somewhere in the remaining degrees of freedom, and this is
accomplished through Breit-Wigner resonances appearing in the
two-meson continuum at points where the mass of the initial
heavy-light meson and light-light resonance are equal.  The single meson
decays have thus achieved the same interpretation as the $\rho$ peak
in the $\pi \pi$ continuum.

	Despite this natural interpretation, it is important to see
explicitly that it supports the correct large $N_c$ counting.  Let us
suppose that a certain class of diagrams gives a naive one-meson weak
decay width of order $N_c^n$ for certain special choices of $M$.  The
equivalent resonant two-meson decay diagram has a naive amplitude of
order $N_c^{(n-1)/2}$ (and hence a naive width $O(N_c^{(n-1)})$), and
includes a propagator of the form\footnote{In the narrow width
approximation, which is automatically satisfied in large $N_c$ QCD, it
does not matter whether $\mu_p$ or $\mu_0$ is chosen as the
coefficient of $\Gamma_p$ in the propagator.  We have found that this
statement is empirically true in our numerical simulations.}
\begin{equation}
\frac{i}{\mu_0^2 - \mu_p^2 + i \mu_p \Gamma_p},
\end{equation}
where the initial meson has mass $\mu_0$, and the resonance, labeled
by $p$, has mass $\mu_p$ and strong width $\Gamma_p = O(1/N_c)$.
Unless $\mu_0^2$ is very close to $\mu_p^2$, the propagator is
$O(N_c^0)$ and the naive large $N_c$ counting is maintained.  However,
when $\mu_0^2 = \mu_p^2$ the previously suppressed factor $\Gamma_p$
becomes dominant and promotes the propagator to a quantity of order
$N_c^1$.  The question becomes, how much area lies under this
Breit-Wigner?  To answer this, we note that the relevant quantity in
the width is the propagator squared.  Since the peak is very tall and
narrow, the rest of the invariant amplitude varies little over the
width of the peak, and thus may be treated as an overall constant.
We may then extend the limits of the integral in $\mu_0^2$ from the
immediate $O(1/N_c)$ neighborhood of $\mu_p^2$ to all values $\mu_0^2
\in (-\infty, +\infty)$.  Noting that
\begin{equation}
\int_{-\infty}^{+\infty} d \mu_0^2 \, \frac{1}{\left( \mu_0^2 -
\mu_p^2 \right)^2 + \mu_p^2 \Gamma_p^2} = \frac{\pi}{\mu_p \Gamma_p} =
O(N_c^1),
\end{equation}
the total weak width becomes $O(N_c^n)$, and we see that the large
$N_c$ counting and unitarity are preserved, exactly as claimed.  More
precisely, if the product $f(\mu_0^2)$ of the invariant amplitude
(except for the propagator), phase space and whatever measure we
choose for the integration over $\mu_0^2$ is a smooth function in the
neighborhood of $\mu_p$, then
\begin{equation}
\int_{-\infty}^{+\infty} d \mu_0^2 \, \frac{f(\mu_0^2)}{\left( \mu_0^2 -
\mu_p^2 \right)^2 + \mu_p^2 \Gamma_p^2}\approx 
\frac{\pi f(\mu_p^2)}{\mu_p \Gamma_p} =
O(N_c^1).
\end{equation}
	It follows that the $N_c$ counting for two-meson decay
diagrams must be modified when it is possible within the topology
class to have a final state consisting of a sole color-singlet $q\bar
q$ pair.  The A and E diagrams, and those subsets of P and PA diagrams
singled out above, satisfy these criteria.  In such cases, the width
is promoted by one power of $N_c$ over the result of naive power
counting.  We obtain finally the large $N_c$ hierarchy for diagram
topologies summarized in Table~1.

\bigskip
\vbox{\medskip
\hfil\vbox{\offinterlineskip
\hrule

\halign{&\vrule#&\strut $\, \,$ \hfil$#$ $\, \,$ \hfil\cr
height0pt&\omit&&\omit&&\omit&&\omit&&\omit&\cr
& N_c^2  && N_c^1 && N_c^0 && N_c^{-1} && N_c^{-2} & \cr
\noalign{\hrule}
& {\rm A}_R && {\rm T} && {\rm P}_R, \, {\rm E}_R && {\rm C}
 && {\rm PA}_R &\cr} \hrule} \hfil
\medskip
\\
\INSERTCAP{1}{Large $N_c$ dependence of two-body meson decay widths
from all quark diagram topologies.  A subscript {\em R} indicates that the
counting is enhanced by $N_c^1$ relative to naive counting by the
presence of resonant intermediate states.}}

	We see that the A diagram actually dominates over the T
diagram studied in Ref.~\cite{GL} by one power of $N_c$, precisely
because resonance intermediates enhance the former and not the latter.
However, in order for both diagrams to appear in a single process,
certain restrictions on the flavors of quarks in the mesons must be
imposed.  This assignment must be performed with some care, because in
the general case one may have to deal with the statistics of identical
mesons.  In order to base our conclusions on as simple a system as
possible, in the present work we have chosen flavors so that such
identical final-state mesons do not occur.  In terms of the labels in
the T diagram of Figs.~6$a$ and the A diagram of 6$b$, we choose
parton 1 to be the heavy quark ``$b$'' of mass $M$, partons 3 and 4 to
refer to identical light quarks, and 2, 3 (= 4), and 5 to be quarks
degenerate with mass $m$ but of different flavors.  In terms perhaps
more familiar, this means that a ``$\bar B$'' meson with flavor
content $(b\bar u)$ decays to two ``pion'' excitations with flavors
$(u' \bar u)$ and $(d \bar u')$, where the $u$, $d$, and hypothetical
$u'$ quarks are degenerate in mass.  Similar assignments may be used
to permit certain topologies and forbid others.

\INSERTFIG{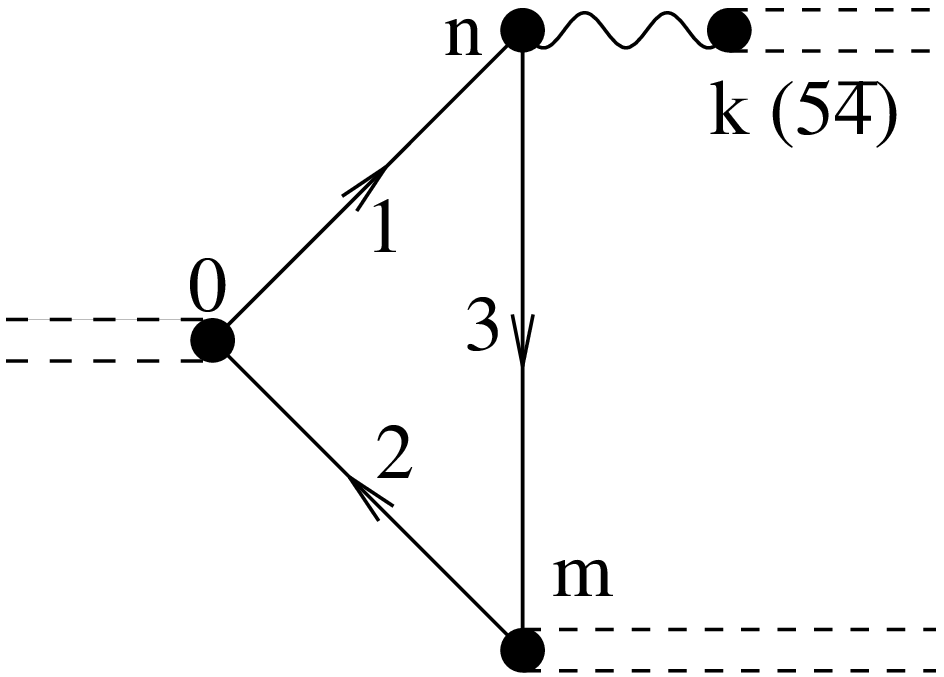}{6$a$}{Diagram for ``tree'' (T) meson exclusive
decay.  Numbers indicate quark labels used in the text (except {\bf 0},
which refers to the ground-state ``$\bar B$'' meson), while letters
indicate the eigenvalue index of meson resonances.  One can also
consider contact-type diagrams, in which the point labeled by\/ {\bf
n} is not coupled to a resonance.}

\INSERTFIG{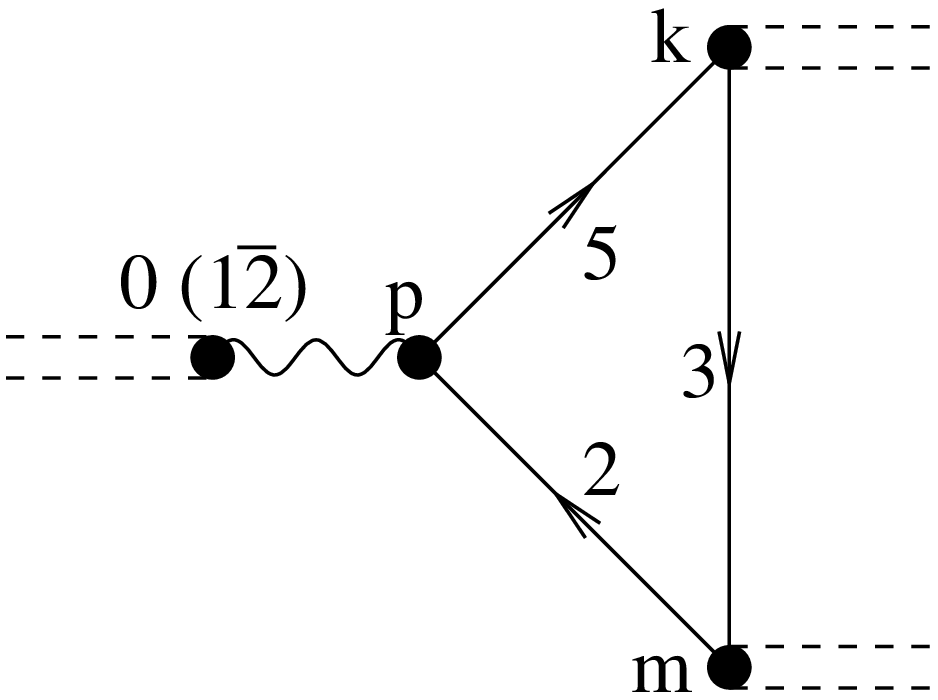}{6$b$}{Diagram for ``annihilation'' (A) meson
exclusive decay.  Numbers indicate quark labels used in the text
(except {\bf 0}, which refers to the ground-state ``$\bar B$'' meson),
while letters indicate the eigenvalue index of meson resonances.}

\section{The 't~Hooft Model and Annihilation Diagrams} \label{review}

	A review of the application of the 't~Hooft model and a
description of the interesting physical peculiarities of 1+1
dimensions appears in Ref.~\cite{GL}.  Here we reprise only those
properties essential to understanding the exceptional features of our
results presented below.

	The 't~Hooft model is defined as SU($N_c)$ Yang-Mills theory
in 1+1 spacetime dimensions with both adjoint (gluon) and fundamental
(quark) degrees of freedom in the limit of infinite group rank $N_c -
1$.  This definition leads to a theory in which those Feynman diagrams
with leading dependence in $N_c$ may be summed explicitly to give
closed-form expressions that are numerically, if not analytically,
soluble.  The archetype of these expressions is the solution for the
two-point irreducible Green function, the 't~Hooft equation:
\begin{equation} \label{tHe}
\mu_n^2 \phi_n^{M\overline{m}} (x) = \left( \frac{M_R^2}{x} +
\frac{m_R^2}{1-x} \right) \phi_n^{M\overline{m}}(x) - \int^1_0 dy \,
\phi_n^{M\overline{m}} (y) \, \Pr \frac{1}{(y-x)^2},
\end{equation}
where $\phi_n$ is the $n$th meson eigenfunction for a quark-antiquark
pair of masses $M$, $m$, while $\mu$ is the meson mass eigenvalue, $x$
is the fraction of the meson momentum carried by the quark in
light-cone coordinates, and $R$ indicates quark mass renormalization:
\begin{equation} \label{mren}
m_a^2 \to m_{a,R}^2 \equiv m_a^2 - g^2 N_c / 2 \pi .
\end{equation}
Note that gauge couplings in 1+1 dimensions have units of (mass)$^1$,
and recall the usual 't~Hooft scaling of the strong coupling $g
\propto 1/\sqrt{N_c}$.  From (\ref{mren}) it then follows that $g^2
N_c / 2 \pi$ serves as the natural unit of mass in the 't~Hooft model,
and this redefinition has already been used in Eq.~(\ref{tHe}).
Indeed, this scale serves much the same purpose in 1+1 as
$\Lambda_{\rm QCD}$ in 3+1, in that heavy quark scaling properties set
in for quark masses a few times $g \sqrt{N_c/2\pi}$\cite{JM,GM1,GM2}.

	Meson wavefunction solutions to the 't~Hooft equation are
either pseudoscalars or scalars, since rotations do not
exist in one spatial dimension except in the residual form of spatial
inversion.

	That three-point 't~Hooft model Green functions may be
expressed in terms of two-point functions was first shown by
Einhorn\cite{Ein}, with various generalizations demonstrated by
Grinstein and Mende\cite{GM1,GM2}.  The method\cite{Mul} of numerically
solving the 't~Hooft equation is discussed in Sec.~\ref{res}, and
described in detail in \cite{GL}.

	The most remarkable characteristic distinct to 1+1 physics
exhibited by these calculations is that two-particle phase space
actually becomes singular as $1/|{\bf p}|$ when the threshold $|{\bf
p}| = 0$ is approached, where $|{\bf p}|$ is the CM spatial momentum
of either outgoing meson.  Nevertheless, these turn out to be rather
weak square root singularities in the heavy quark mass $M$ and
therefore have pronounced effects on the width only at points over a
relatively small measure in $M$.  The corresponding familiar 3+1
expression, on the other hand, is proportional to $|{\bf p}|^1$ and
thus vanishes at threshold.

	Of much greater numerical relevance in any number of
dimensions are the Breit-Wigner lineshapes produced by resonant
intermediate states.  As discussed in Sec.~\ref{diags}, such
resonances enhance the width by a full factor of $N_c^1$ for the A but
not the T diagram, and therefore were not an issue in Ref.~\cite{GL}.
Since explicit factors of $N_c$ no longer appear simply as a single
overall coefficient in the width, one must choose a particular value
for $N_c$ in order to obtain numerical results.  Quark-hadron duality
should then be studied by taking $N_c$ as large as numerically
feasible; we return to a discussion of this point in Sec.~\ref{res}.

\section{Partonic Widths} \label{inc}

	The computation of the inclusive width for the decay of a
heavy-light meson via the annihilation diagram at the Born level is
fairly straightforward, and may be accomplished through two
complementary approaches, each of which has its own advantages.

	The quark Hamiltonian used for this calculation is given by
\begin{equation}
{\cal H} = G \Delta_{\mu\nu} \left[ \bar q_2 \gamma^\mu \left( c_V +
c_A \gamma_5 \right) b \right] \left[ \bar q_5 \gamma^\nu \left( c_V +
c_A \gamma_5 \right) q_2 \right] ,
\end{equation}
where
\begin{equation}
\Delta_{\mu\nu} = g_{\mu\nu} + \xi (p) p_\mu p_\nu
\end{equation}
is the tensor structure of the weak interaction propagator with
momentum transfer $p$.  If one adopts notation analogous to that of
the Standard Model, then in unitary gauge, such that charged Higgs
bosons become irrelevant,
\begin{equation}
\xi = -1/M_W^2 .
\end{equation}
Similarly,
\begin{equation}
G = 2\sqrt{2} G_F \frac{M_W^2}{M_W^2 - p^2} V_{21} V^*_{25} ,
\end{equation}
where $G_F/\sqrt{2} = g_2^2 / 8M_W^2$ as usual.  To complete the
analogy, in the usual $V-A$ theory one would have $c_V = - c_A = 1/2$,
but we leave these as free parameters.

	The partonic calculation of the width must incorporate the
annihilation of the initial meson as well as the production of final-state
free quarks.  Rather than decomposing the initial meson into a
quark-antiquark pair free but correlated in such a way as to guarantee
the desired total quantum numbers, we include the full meson coupling
through
\begin{equation}
\label{eq:decayconstant}
\left< 0 \left| c_V V^\mu + c_A A^\mu \right| n \right> = \left( c_V
\epsilon^{\mu \nu} + c_A g^{\mu \nu} \right) f_n  p_\nu ,
\end{equation}
where $V^\mu = \bar q \gamma^\mu q$, $A^\mu = \bar q \gamma^\mu
\gamma_5 q = \epsilon^{\mu\nu} V_\nu$, $\epsilon_{01} = +1$ and $f_n$
is the decay constant of meson $n$.  In the 't~Hooft model
\begin{equation} \label{fdef}
f_n = \sqrt{\frac{N_c}{\pi}} \, c_n \equiv \sqrt{\frac{N_c}{\pi}}
\int_0^1 dx \, \phi_n (x) .
\end{equation}

	The first computational approach evaluates the amplitude
directly obtained from the diagram Fig.~3, where the quark-antiquark
pair created by the weak current is represented by free spinors, and the
on-shell process has $p^2 = \mu_0^2$.  Using the Dirac equation and
the 1+1 identity $\gamma^\mu \gamma_5 = \epsilon^{\mu\nu} \gamma_\nu$,
one obtains
\begin{equation} \label{amp}
{\cal M}_A = -2\sqrt{\frac{N_c}{\pi}} G c_0 m \left[ c_V^2 - c_A^2
\left( 1 + \xi \mu_0^2 \right) \right] \left[ \bar u_5 \gamma_5 v_2
\right] .
\end{equation}
The width is given by
\begin{equation} \label{ph}
\Gamma = \frac{1}{4\mu_0^2 |{\bf p}|} \left| {\cal M}_A \right|^2 .
\end{equation}
Including a factor $N_c$ for outgoing quarks and $1/\sqrt{N_c}$ for
normalizing them into a color singlet, one obtains
\begin{equation} \label{irate}
\Gamma_{\rm part} = N_c^2 G^2 \left[ c_V^2 - c_A^2 \left( 1 + \xi
\mu_0^2 \right) \right]^2 \frac{2m^2 c_0^2}{\pi \sqrt{\mu_0^2 -
4m^2}}.
\end{equation}

	It is interesting to note that the rate vanishes in the limit
of massless light quarks.  This follows trivially from the observation
that in the $m\to0$ chiral limit both vector and axial-vector currents
(of light quarks) are conserved.  Since for large $N_c$ the A
amplitude factorizes, one must contract the vector index in
Eq.~(\ref{eq:decayconstant}) with that of the current producing the quarks.
Contraction with $p_\mu$ corresponds to taking the divergence of the
currents, while contraction with $\epsilon_{\mu\nu}p^\nu$ corresponds
to first exchanging the role of vector and axial-vector currents, and
then taking the divergence.  Incidentally, this argument also applies
in 3+1 dimensions.

	Since the decay constant $f_0 \propto c_0$ scales as
$1/\sqrt{M}$\cite{GM1} and $\mu_0 \propto M$ as $M \to \infty$, the
asymptotic $M$ behavior of the annihilation diagram width $\Gamma_{\rm
part}$ is $1/M^2$.  This is to be contrasted with the tree diagram
asymptotic width\cite{GL}, which grows as $M^1$.

	The second computational approach evaluates the width by
calculating the loop integral in Fig.~4$b$ and then using unitarity to
cut the diagram and reveal the on-shell result.  Assuming that each
quark has a nonzero value of some conserved quantum number such as
electric charge, at most one of the two diagrams in Fig.~4$b$ can
occur in a given physical process.  In the present case, it is the
first diagram, which has external quark-antiquark pair 1,2 and
internal quark-antiquark pair 5,2.

	Standard techniques show that the inclusive width to any final
states $X$ in $D$ spacetime dimensions is given by
\begin{equation} \label{genrate}
\Gamma (\bar B \to X) = \frac{1}{M} \, {\rm Im} \, i \! \int d^D x \,
\left< \bar B \left| T {\cal H}^\dagger (x) {\cal H} (0) \right| \bar
B \right> ,
\end{equation}
so that one requires only (one-half of) the discontinuity in the
imaginary part of the loop diagram, which begins at values of energy
where on-shell
intermediate states appear.  The factorized four-quark operator
in ${\cal H}$ is used to annihilate and create a $\bar B$ meson, as is
again quantified by Eq.~(\ref{eq:decayconstant}), leaving a vacuum
amplitude of the product of two currents.  Such factorization is a
consequence of large $N_c$.  Retaining only the internal
quarks and their couplings to the weak current, one obtains for the
diagram
\begin{eqnarray} \label{aloop}
& & \left( -\frac{iN_c}{\pi} \right)
\left( c_V g^\mu{}_\rho   + c_A \epsilon^\mu{}_\rho   \right)
\left( c_V g^\nu{}_\sigma + c_A \epsilon^\nu{}_\sigma \right)
\left( g^{\rho\sigma} - p^\rho p^\sigma/p^2 \right) \nonumber \\ & &
\hspace{4em} \cdot \left[ 1 - \frac{4m^2}{p^2} \frac{1}{\sqrt{4m^2/p^2
-1}} \tan^{-1} \frac{1}{\sqrt{4m^2/p^2 -1}} \right].
\end{eqnarray}
The discontinuity of the bracketed quantity across the cut for $p^2
\ge 4m^2$ is given by
\begin{equation} \label{disc}
\frac{4m^2}{p^2} \frac{1}{\sqrt{1 - 4m^2/p^2}}.
\end{equation}
The external couplings and weak current propagators are given by
\begin{equation} \label{excoup}
-G^2 f_0^2 p^\tau p^\omega \bigl[ c_V \epsilon^\kappa{}_\tau + c_A
g^\kappa{}_\tau \bigr] \bigl[ c_V \epsilon^\lambda{}_\omega + c_A
g^\lambda{}_\omega \bigr] \left[ g_{\mu\kappa} + \xi p_\mu
p_\kappa \right] \left[ g_{\nu\lambda} + \xi p_\nu p_\lambda \right] .
\end{equation}
Substituting into (\ref{genrate}) the expression (\ref{excoup})
contracted with (\ref{aloop}) using the discontinuity in (\ref{disc}),
and finally replacing $f_0$ using (\ref{fdef}), one again obtains the
partonic rate (\ref{irate}).  In all subsequent expressions, we take
$\xi = 0$, corresponding to $M_W \to \infty$.

	Once this computation is phrased in terms of the vacuum
amplitude of the product of two currents, one is tempted to replace
the product of two currents by an OPE.\@ However, this procedure is
poorly justified, if at all, since the momentum across the currents,
$p$, is neither in the deep Euclidean region (where the OPE is
systematic) nor is it integrated over a region in such a way that the
contour of integration can be deformed so that it lies (mostly) in the
deep Euclidean region.  The physical quantity of interest is the rate
at a given heavy meson mass, so that $p^2=\mu_0^2$ is timelike.  One
can, however, consider integrating the rate over the variable $p^2$.
Then, as in the more familiar case of $e^+e^-\to{\rm hadrons}$, the
contour can be deformed and the integral is dominated by the leading
term in the OPE.\@ Putting aside the question of physical utility of
this exercise (in reality, unfortunately, we cannot vary the mass of
the $\bar B$ meson), in the 't~Hooft model it has long been
known\cite{CCG,Ein} that this leading order OPE result is reproduced
by the sum over intermediate resonant states; for clarity, we
demonstrate this result in the current notation below.  Large $N_c$
counting dictates that only single intermediate states contribute.  If
global duality (that is, including integration over $p^2$) is operative in the
rate for annihilation decays, it must be through some nontrivial
interplay between these resonances and the inclusion of widths for
internal meson propagators in the exclusive rates, as discussed in
Sec.~\ref{diags}.

	In the 't~Hooft model one may explicitly check duality for the
vacuum amplitude of the product of two currents in
the limit of $p^2$ large and in any complex direction except along the
positive real axis (where meson poles occur). In fact, this limit was
considered first by
Callan, Coote, and Gross\cite{CCG}, with a number of refinements by
Einhorn\cite{Ein}, but it is instructive to see how the calculation
proceeds when arbitrary combinations of vectorlike currents are
included.  Defining $\tilde p^\mu = \epsilon^{\mu\nu} p_\nu$ and using
$\epsilon^{\mu\kappa} \epsilon^\nu{}_\kappa = -g^{\mu\nu}$, the $p^2 \gg
m^2$ limit of the loop expression (\ref{aloop}) reads
\begin{equation} \label{du1}
\left( -\frac{iN_c}{\pi} \right) \left\{ \left( c_V^2 - c_A^2 \right)
g^{\mu\nu} - \frac{1}{p^2} \left( c_V^2 p^\mu p^\nu + c_A^2 \tilde
p^\mu \tilde p^\nu \right) - \frac{c_V c_A}{p^2} \left( p^\mu \tilde
p^\nu + \tilde p^\mu p^\nu \right) \right\} .
\end{equation}
One the other hand, the hadronic vertex is defined by
Eq.~(\ref{eq:decayconstant}), and so the loop written in terms of
resonance contributions reads
\begin{equation} \label{pdu2}
\left(c_A \epsilon^\mu{}_\tau + c_A g^\mu{}_\tau \right) \left( c_V
\epsilon^\nu{}_\omega + c_A g^\nu{}_\omega \right) p^\tau p^\omega
\sum_n \frac{if_n^2}{p^2-\mu_n^2} .
\end{equation}
As $p^2 \to$ complex $\infty$, the sum
becomes
\begin{equation}
\frac{i}{p^2} \sum_n f_n^2 = \frac{iN_c}{\pi p^2} \sum_n c_n^2 =
\frac{iN_c}{\pi p^2} ,
\end{equation}
which uses the definition (\ref{fdef}) and the
completeness relation
\begin{equation}
\sum_n \phi_n (x) \phi_n (y) = \delta (x-y).
\end{equation}
In this limit, (\ref{pdu2}) contracts to
\begin{equation} \label{du2}
\left( + \frac{iN_c}{\pi} \right) \left\{ c_V^2 \frac{\tilde p^\mu
\tilde p^\nu}{p^2} + c_A^2 \frac{p^\mu p^\nu}{p^2} + \frac{c_V
c_A}{p^2} \left( \tilde p^\mu p^\nu + p^\mu \tilde p^\nu \right)
\right\} .
\end{equation}
To see that (\ref{du1}) and (\ref{du2}) are equal requires one to
recognize the following (equivalent) tensor identities, which hold in
1+1:
\begin{eqnarray}
p^\mu p^\nu - p^2 g^{\mu\nu} & = & \tilde p^\mu \tilde p^\nu , \nonumber
\\
\tilde p^\mu \tilde p^\nu + p^2 g^{\mu\nu} & = & p^\mu p^\nu .
\end{eqnarray}

	It goes without saying that a demonstration of the validity of
an OPE for hadronic widths in the real world of four dimensions and
three colors would be much more subtle, as one loses some simplifying
elements such as factorization.

	If it nevertheless can be shown that the nonleptonic expansion
admits a well-defined OPE, then the diagrams of Fig.~4$b$ enter as
effective four-quark operators of the form
\begin{equation}
{\cal O}_A = \bar b \Gamma^\mu q_2 \, \bar q_2 \Gamma_\mu b ,
\end{equation}
where $\Gamma^\mu$ represents the vectorlike ($V^\mu$ and $A^\mu$)
Lorentz structures.  In contrast, the T diagram arises from cutting
the loop diagram of Fig.~4$a$, and enters the effective OPE through
the leading operator ${\cal O}_T = \bar b b$.  Since fermion fields in
$D$ spacetime dimensions have engineering dimension $M^{(D-1)/2}$, by
naive power counting the A diagram width in 1+1 might be expected to
be only $1/M$ suppressed compared to that of the T diagram.
Schematically, the OPE-like expression for the width reads
\begin{equation} \label{ope}
\Gamma (\bar B \to X) \sim G_F^2 M^{2D-4} \left\{ \left< \bar B \left|
\bar b b \right| \bar B \right> + \cdots + \frac{1}{M^{D-1}} \left<
\bar B \left| \bar b \Gamma^\mu q_2 \, \bar q_2 \Gamma_\mu b \right|
\bar B \right> + \cdots \right\} ,
\end{equation}
where the ellipses indicate subleading terms for both the T and A
contributions, and numerous overall coefficients as well as
perturbative short-distance corrections have been suppressed for
simplicity.  The overall mass factor is obtained by noting that the
mass dimension of $G_F$ is $M^{2-D}$, while the $\bar B$ bra and ket
are normalized to $2M$ particles per unit volume and thus have mass
dimension $M^{(1-D/2)}$.  Each term in the braces has dimension $M^1$.

	However, this reasoning does not take into account the light
quark mass suppression induced by taking the divergence of the light
quark current.  One obtains an additional suppression $m^2/M^2$ as
discussed above, and thus (\ref{ope}) does indeed predict $\Gamma_A
\propto 1/M^2$ as $M \to \infty$ in 1+1, in agreement with
(\ref{irate}).

\INSERTFIG{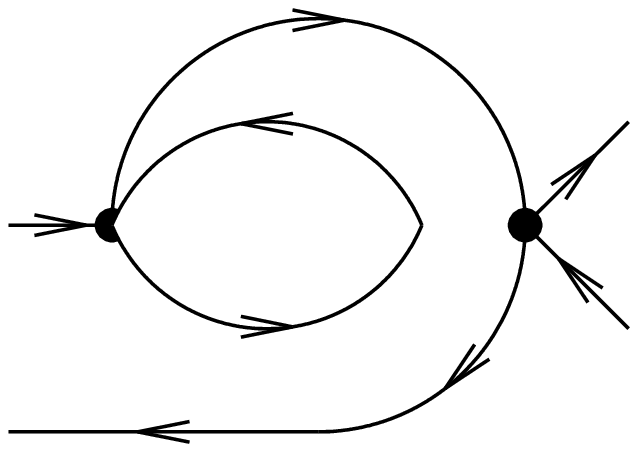}{7}{Feynman diagram topology for the
interference between A and T amplitudes.}

	An important point to note is the absence of interference
effects between A and T amplitudes in the total width predicted by the
OPE-like expansion, Eq.~(\ref{ope}).  The diagram in Fig.~7 would
contribute to such an interference effect, but is of order $N_c^1$
rather than $N_c^{3/2}$ as suggested by Table~1. The absence of
interference in the OPE method is not by itself proof of failure of
the method since there could be cancellations among the exclusive
channels that effectively cancel the interference effects.  

\section{Hadronic Widths in the 't~Hooft Model} \label{exc}

	In Ref.~\cite{GL} the amplitude for each allowed exclusive
channel proceeding through the T diagram of Fig.~6$a$ was computed in
terms of various sums and overlaps of 't~Hooft model wavefunctions.
Such an expression, Eq.~(5.8) or (5.18) in that work, represents the
exact nonperturbative calculation of the invariant amplitude within
the 't~Hooft universe of 1+1 spacetime dimensions and large $N_c$.

        The analogous calculation for the A diagram of Fig.~6$b$ is
almost identical, and essentially amounts to a reassignment of
indices, as expected from crossing symmetry.\footnote{Since $\mu_0^2 >
(\mu_k + \mu_m)^2$ for all on-shell processes, it follows that
$\omega$ defined here in (\ref{om}) always lies in $[0,1]$, and one
may directly use the analogue of (5.8) in \cite{GL} rather than
worrying about ``backsolving'' or ``contact terms'' as described in
the previous work.}  The expressions in terms of form factors are very
similar to those for the T diagram, and so we instead present only the
result for the invariant amplitude.  As before, {\bf 0} refers to the
initial ground state ($1 \bar 2$) meson, and is now directly coupled
to the flavor-changing current.  The light meson at the other end of
this current, labeled by {\bf p}, has quantum numbers ($5 \bar 2$);
{\bf m} as before refers to the final-state ($3 \bar 2$) meson, and
{\bf k} (which is no longer coupled to the flavor-changing current),
has quantum numbers ($5 \bar 3$): Recall the statement in
Sec.~\ref{diags} that, in order to have both T and A diagrams, we
require quarks 3 and 4 to be identical.  The kinematic variable, now
defined by $\omega \equiv q_- / p_-$, is given for this diagram by
\begin{equation} \label{om}
\omega (p^2) = \frac{1}{2} \left[ 1 + \left( \frac{\mu_k^2 -
\mu_m^2}{p^2} \right) - \sqrt{1- 2 \left( \frac{\mu_k^2 +
\mu_m^2}{p^2} \right) + \left( \frac{\mu_k^2 - \mu_m^2}{p^2}
\right)^2} \, \right].
\end{equation}
Here we see that the relevant threshold is that of {\bf 0}$
\rightarrow${\bf mk}, {\it i.e.}, $p^2 = (\mu_k + \mu_m)^2$.
The invariant amplitude for states above this threshold is given by
\begin{eqnarray} \label{ma}
{\cal M}_A & = & G c_0 \sqrt{\frac{N_c}{\pi}}
\sum_p \Biggl[ \left[ (c_V^2-c_A^2) \left( 1 + (-1)^p \right) \right]
-\xi p^2 c_A \left[ (c_V + c_A) (-1)^p - (c_V - c_A) \right] \Biggr]
\nonumber \\
& & \hspace{6em} \cdot \frac{c_p \mu_p^2}{(p^2 - \mu_p^2 + i \mu_p
\Gamma_p)} F_{pkm} (\omega_0) ,
\end{eqnarray}
where now the on-shell process has $p^2 = \mu_0^2$, $\omega_0 \equiv
\omega(p^2 = \mu_0^2)$, and the triple overlap is given by
\begin{eqnarray} \label{fpkm}
\lefteqn{F_{pkm} (\omega) \equiv} & & \nonumber \\
& & \left[ \frac{1}{1-\omega} \int_0^\omega dv \,
\phi_p^{5\bar 2} (v) \phi_k^{5\bar 3} \left(\frac v \omega
\right) \Phi_m^{3 \bar 2} \left( \frac{v-\omega}{1-\omega} \right) -
\frac 1 \omega \int^1_\omega dv \, \phi_p^{5\bar 2} (v)
\Phi_k^{5\bar 3} \left( \frac v \omega \right) \phi_m^{3\bar 2}
\left( \frac{v-\omega}{1-\omega} \right) \right] ,
\end{eqnarray}
where the meson-quark vertex function is defined by
\begin{equation}
\Phi_n^{M\overline{m}} (z) = \int_0^1 dy \, \phi_n^{M\overline{m}} (y)
\, \Pr \frac{1}{(y-z)^2} .
\end{equation}
Note also the presence of the partial width $\Gamma_p$ for light-light
meson {\bf p} strong decay into mesons {\bf k} and {\bf m}.  As argued
in the Sec.~\ref{diags}, such an inclusion is essential to give a
consistent large $N_c$ power counting for the two-meson weak decay
diagram.  Explicitly,
\begin{equation} \label{swidth}
\Gamma_p (\omega_p) = \frac{2\pi}{N_c \mu_p} \sum_{k,m}
\left[(\mu_p^2)^2 - 2 \mu_p^2 \left( \mu_k^2 + \mu_m^2 \right) +
\left( \mu_k^2 - \mu_m^2 \right)^2 \right]^{-1/2} |F_{pkm}
(\omega_p)|^2,
\end{equation}
where, using (\ref{om}), $\omega_p \equiv \omega (p^2 = \mu_p^2)$.

	The total rate $\Gamma_{\rm had}$ for the A diagram is then
obtained by squaring (\ref{ma}), inserting the result into (\ref{ph}),
and summing over all allowed final states {\bf k} and {\bf m}.  This
value is to be compared with the Born term expression (\ref{irate})
as a test of quark-hadron duality.

	One may also consider direct numerical comparisons between the
T diagram total hadronic rate exhibited in Fig.~4 of \cite{GL} and
$\Gamma_{\rm had}$ computed here; the correct  procedure to follow
in this case is somewhat ambiguous, since the former rate is of order
$N_c^1$, while the latter integrates to order $N_c^2$ when resonant
contributions are taken into account.  Therefore, if one works only in
the strict large $N_c$ limit, the T diagram is infinitely small
compared to the A diagram.  However, while the 't~Hooft model is of
course only exactly true when $N_c \to \infty$, this limit is believed
to survive the inclusion of $O(1/N_c)$ corrections\cite{CCG}.
Moreover, since numerous studies in the literature show that the
phenomenological predictions of the large $N_c$ expansion survive even
for $N_c$ as small as 3, we suppose that a quantitative comparison
between the T and A widths has merit even for small $N_c$.

\section{Results and Discussion} \label{res}

The 't~Hooft equation is solved numerically by means of the Multhopp
technique\cite{Mul}, by which the integral expression (\ref{tHe})
is converted to
an equivalent eigenvector equation amenable to solution using
computers. The results presented here were computed with a basis set
of $K=200$ eigenfunctions. As a check that this set is sufficiently
large, the computation was repeated using of $K=50$. Figure 8 shows the
ratio of the total width, solely from the annihilation topology,
computed for $K=50$ to that for $K=200$, in the case of ($a$)
$N_c=10$, and ($b$) $N_c=1$. In both cases the ratio of the
Gaussian-smeared widths [see below, Eq.~(\ref{eq:smearing-discrete})]
is also shown, with a Gaussian width of ($a$) $\Delta M=1.2$ and
($b$) $\Delta M=0.4$, in mass units of $g\sqrt{N_c/2\pi}$. In either
case the difference for widths is
never more than 30\%, while for the average width it is never more
than 10\%, and less than 5\% in the region $M>10$.

\def\epsfsize#1#2{0.38#2}
\INSERTtwoFIGs{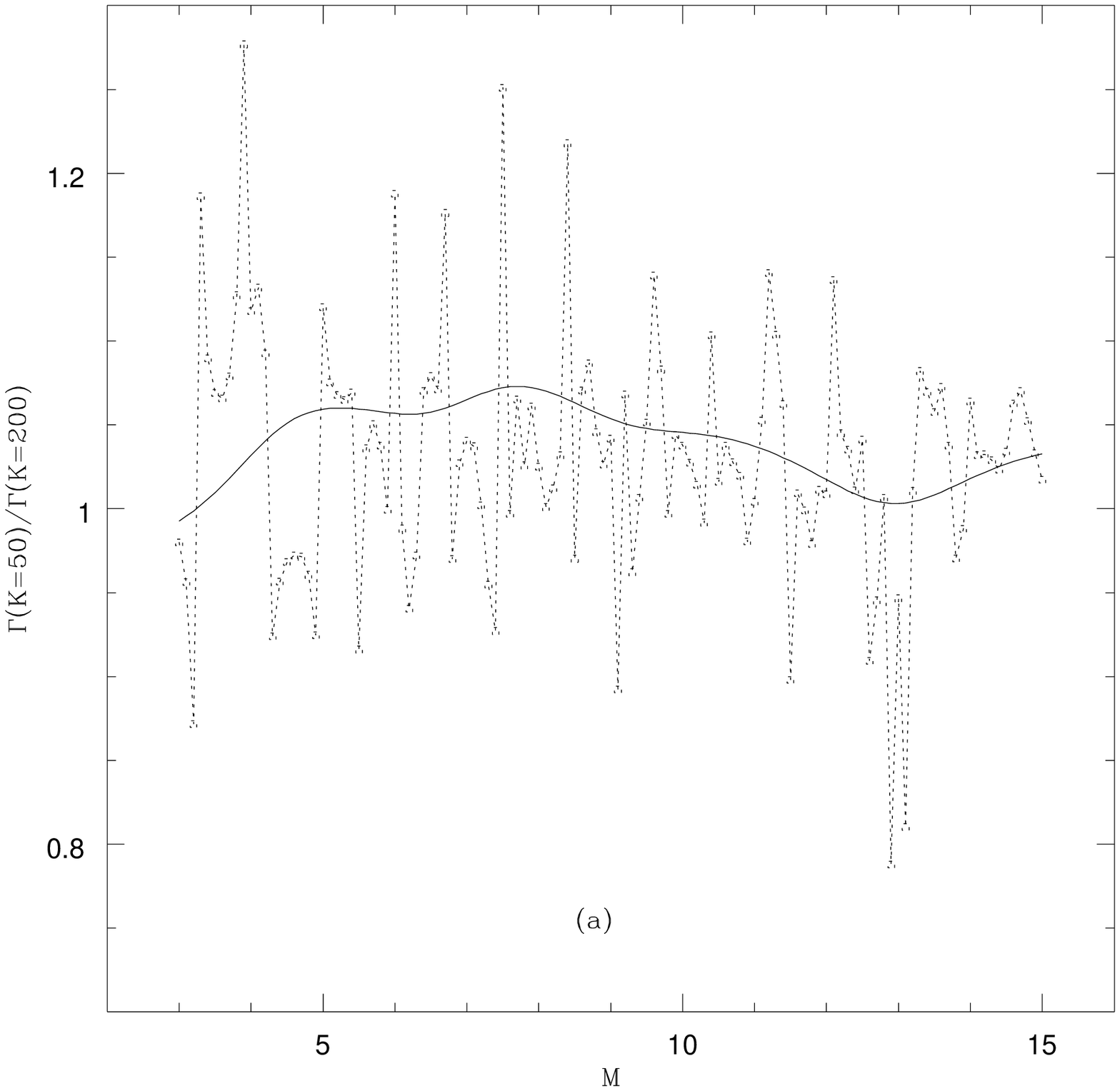}
{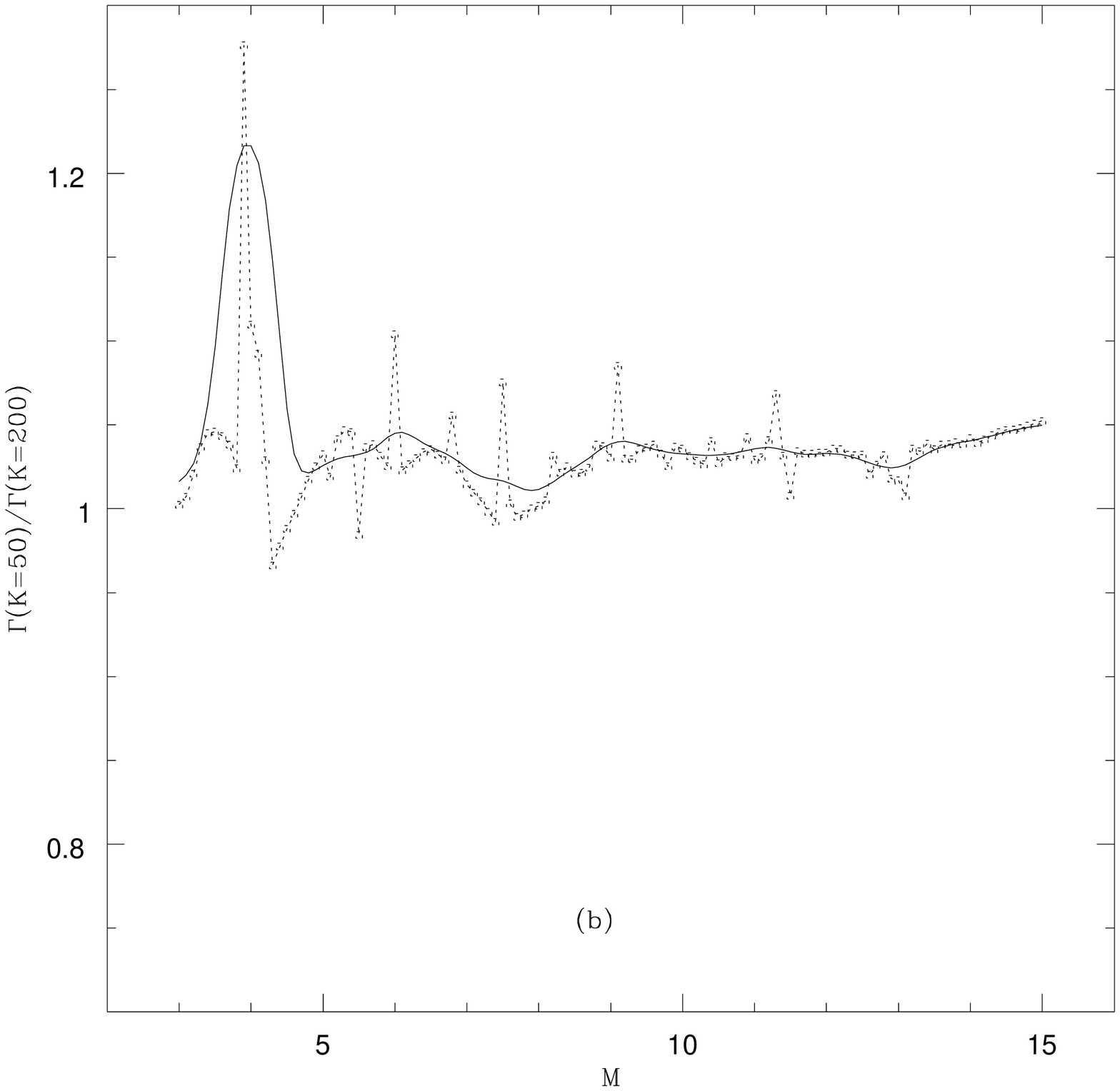}{8}{Ratio of the total width, solely
from the annihilation topology, computed for $K=50$ to that for
$K=200$, in the case of ($a$) $N_c=10$, and ($b$) $N_c=1$. In both
cases the ratio of the Gaussian-smeared widths is also shown, with a
Gaussian width of  $\Delta M=1.2$ in ($a$) and  $\Delta M=0.4$ in ($b$).
The unit of mass here and in all subsequent figures is $g\sqrt{N_c/2\pi}$.}

	In all calculations we choose a single fixed value of mass
common to all the light quarks, $m=0.56$. The range of $M$ over which
calculation of $\Gamma_{\rm had} (M)$ is feasible is limited primarily
by the rapidly increasing number of exclusive channels open to the
decay of the $\bar B$ meson as $M$ increases, and the concomitant
computing time required for the necessary integral overlaps.  In
practice, we limit our studies to the range from $M = 2.28$ (the
lightest heavy quark mass that creates a $\bar B$ with just enough
mass to decay to two ground-state light-light mesons) to $M=15.00$ (at
which point almost 150 exclusive channels are open) in units of $g
\sqrt{N_c/2\pi}$.

	Incidentally, it is known that the standard Multhopp technique
leads to inaccurate 't~Hooft wavefunctions when $m \ll 1$.  In
Ref.\cite{Brow}, an improved version of the Multhopp method is
developed, which does a much better job calculating the wavefunctions
near $x=0$ and 1 when $m$ is small.  One may question whether $m=0.56$
is large enough that our results, computed by the standard technique,
are reliable.  We find that computing with the technique of
\cite{Brow} tends to change the results by only a few parts in $10^4$.
Therefore, we are confident in presenting numerical results below that
use the standard Multhopp technique.

The Breit-Wigner resonances, which become infinitely tall and
narrow as $N_c \to \infty$, present much more severe singularities in
$\Gamma_{\rm had}$ than the phase space singularities discussed in
Sec.~\ref{review}.  Indeed, without proper regularization they are
non-integrable, and therefore no amount of  averaging, or
``smearing,'' could produce a finite result.  One must include the
$O(1/N_c)$ strong widths of the meson resonances coupled to the weak
current, as discussed in Sec.~\ref{diags}.

	The strong widths are interesting in their own right.  The
exact expression for the widths is given in Eq.~(\ref{swidth}).  In
Fig.~9 we show the results of this calculation for meson excitation
numbers from 0 to 155, along with a fit function
\begin{equation} \label{sfit}
\Gamma^{\, \rm fit}_n = 22 \times \frac{0.44}{\pi^2 N_c} \sqrt{n-1} .
\end{equation}

\def\epsfsize#1#2{0.70#2}
\INSERTFIG{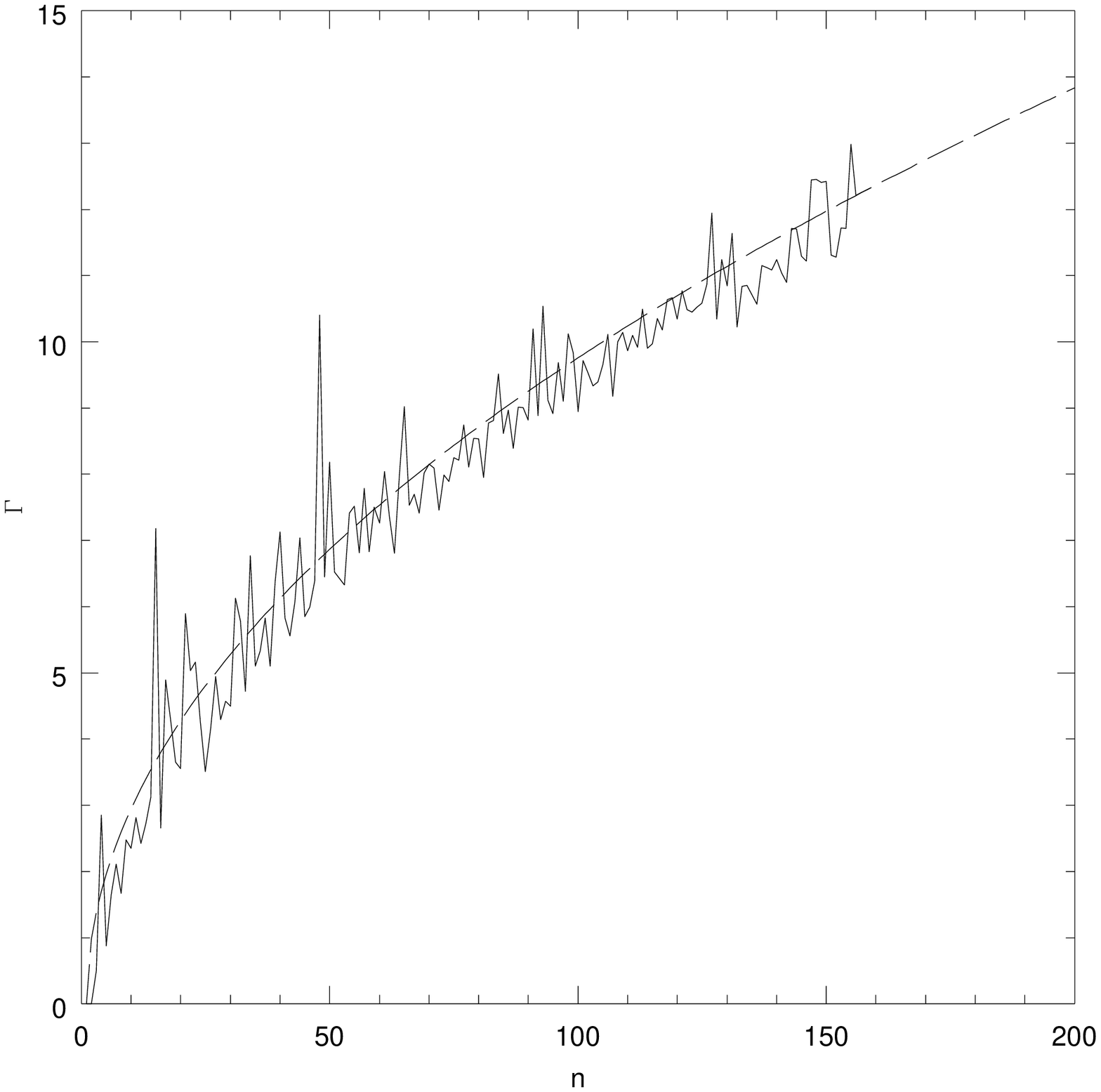}{9}{Meson strong decay widths in units of
$g \sqrt{N_c/2\pi} \cdot 1/N_c$ as computed via
(\ref{swidth}).  The smooth dashed curve is the fit function
(\ref{sfit}).}

The coefficient is written in this peculiar format in order to
facilitate comparison with the results of Blok {\it et al.}\cite{Blok},
whose best fit appears to be a factor 22 smaller than ours.  There are
two differences between their calculation and ours.  First, they
work with massless light quarks $m=0$ so that they may use analytic
expressions for 't~Hooft wavefunctions.  Second, their light quarks,
unlike ours, appear to be chosen as indistinguishable, leading to
a different pattern of interference terms in the invariant
amplitudes.  When strong widths for $n > 155$ are needed, we use the
fit function $\Gamma^{\, \rm fit}_n$.

	In the present case in which strong widths can become
important, factors of $N_c$ no longer appear simply as an overall
coefficient, as was the case in \cite{GL}.  We are thus forced to
choose particular values of $N_c$ in order to obtain numerical
results.  The ideal, of course, is to choose $N_c$ as large as
possible in order to approach the results of the exact 't~Hooft model.
However, such a choice makes the Breit-Wigners taller, narrower, and
thus harder to average over $M$, so that obtaining a
smooth result for comparison with the one computed perturbatively is
more difficult.  In particular, one is forced to choose the range
$\Delta M$ over which the averaging function has support to be larger
and larger in order to obtain at last a smooth result for~$\Gamma_{\, \rm
had}^{\, \rm avg}$.

	The  averaging, or ``smearing,'' of $\Gamma_{\, \rm
had} (M) \to \Gamma_{\, \rm had}^{\, \rm avg} (M)$ in the heavy quark mass
$M$ is carried out here by multiplying $\Gamma_{\rm had} (M)$ at a
series of points $M_0$ over the range of $M$ by a suitably chosen
smearing function, and then normalizing by the area under this
function.  In practice we use a Gaussian of width $\Delta M/\sqrt2$:
\begin{equation}
\label{eq:smearing-discrete}
\Gamma_{\, \rm had}^{\, \rm avg} (M) = \frac{\sum_{M_0} \exp
\left[ -\frac{(M-M_0)^2}{(\Delta M)^2} \right] \Gamma_{\rm had}
(M)}{\sum_{M_0} \exp \left[ -\frac{(M-M_0)^2}{(\Delta M)^2}
\right]} ,
\end{equation}
and the points $M$ and $M_0$ are chosen at intervals of 0.1 mass
units.  It should be pointed out that this smearing produces a small
spurious result at the edges of the fitting range if the function to
be smeared has a nonzero derivative (apart from noise in the function)
at these points.  For example, suppose one smears a linearly
decreasing function near its endpoint.  The smearing function of
course samples only points to the left of the endpoint, where the
function is uniformly larger than it is at the endpoint itself, and so
the smeared result is slightly higher than expected.  One of many
cures is to compare two curves smeared in the same way, which produces
the same spurious effect in both, and so such curves may be compared
directly.

\INSERTFIG{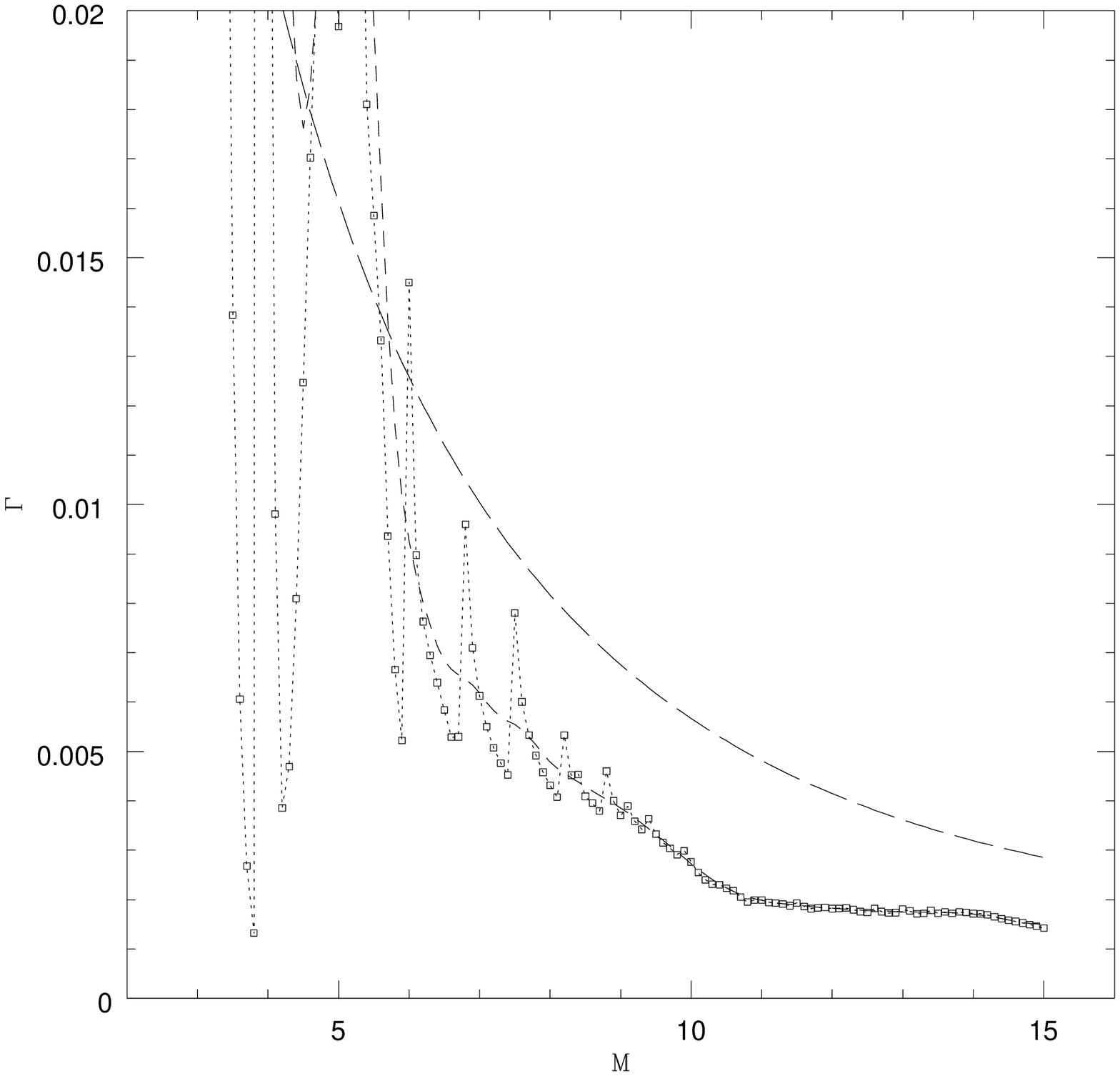}{10}{Total annihilation width
of the ``$\bar B$'' meson as a function of the heavy quark mass
$M$, for  $N_c=1$ (dotted line). The smeared
width defined in Eq.~(\ref{eq:smearing-discrete}) is shown
in short dashes, and was computed using $\Delta M=0.4$. The
partonic width is shown in long dashes.  Units of the widths here and
below are $N_c^1 G^2 (c_V^2 - c_A^2)^2$.}

\INSERTFIG{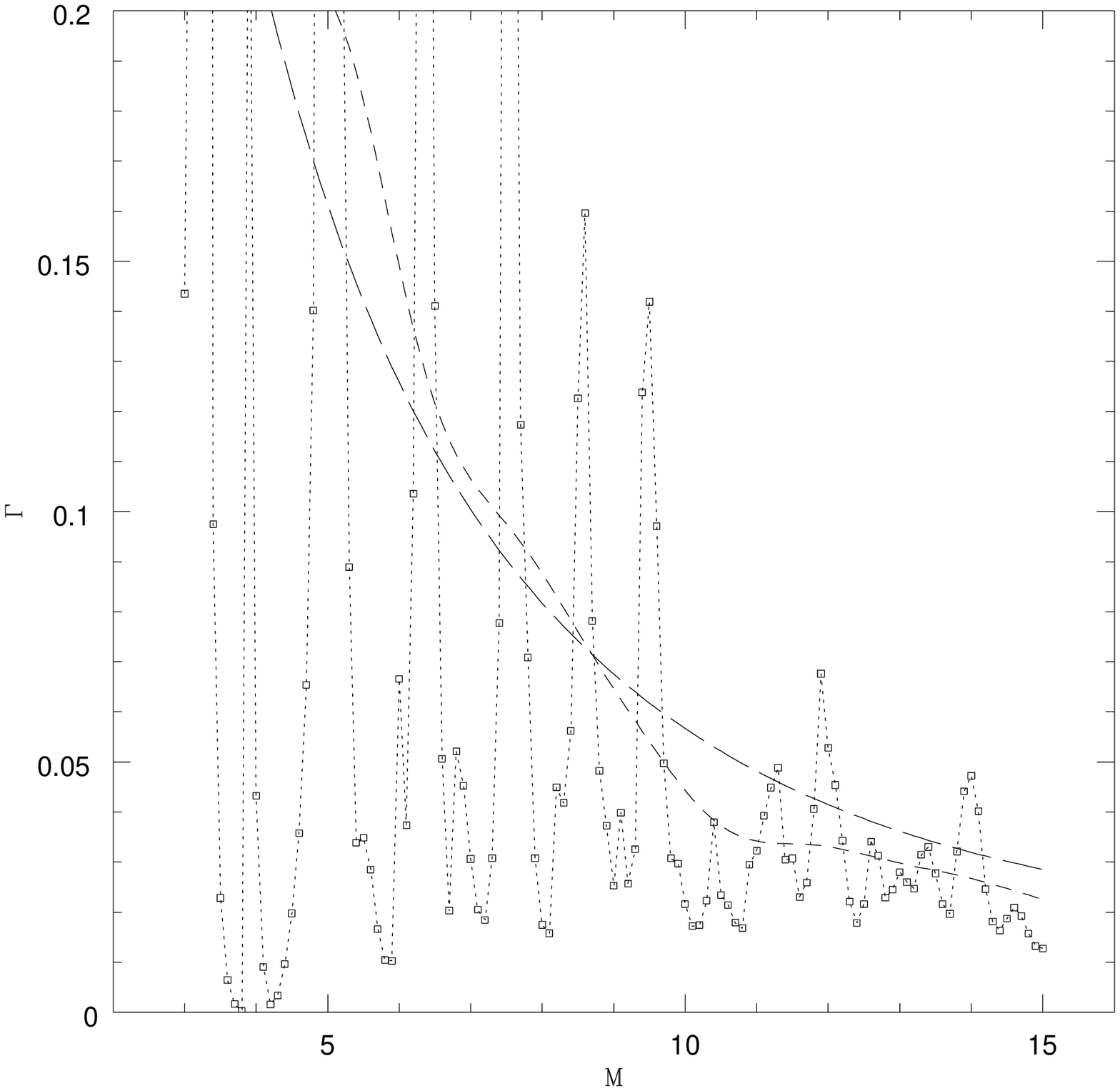}{11}{As in Fig.~10, but with
$N_c=10$ and $\Delta M=1.2$.}

\INSERTFIG{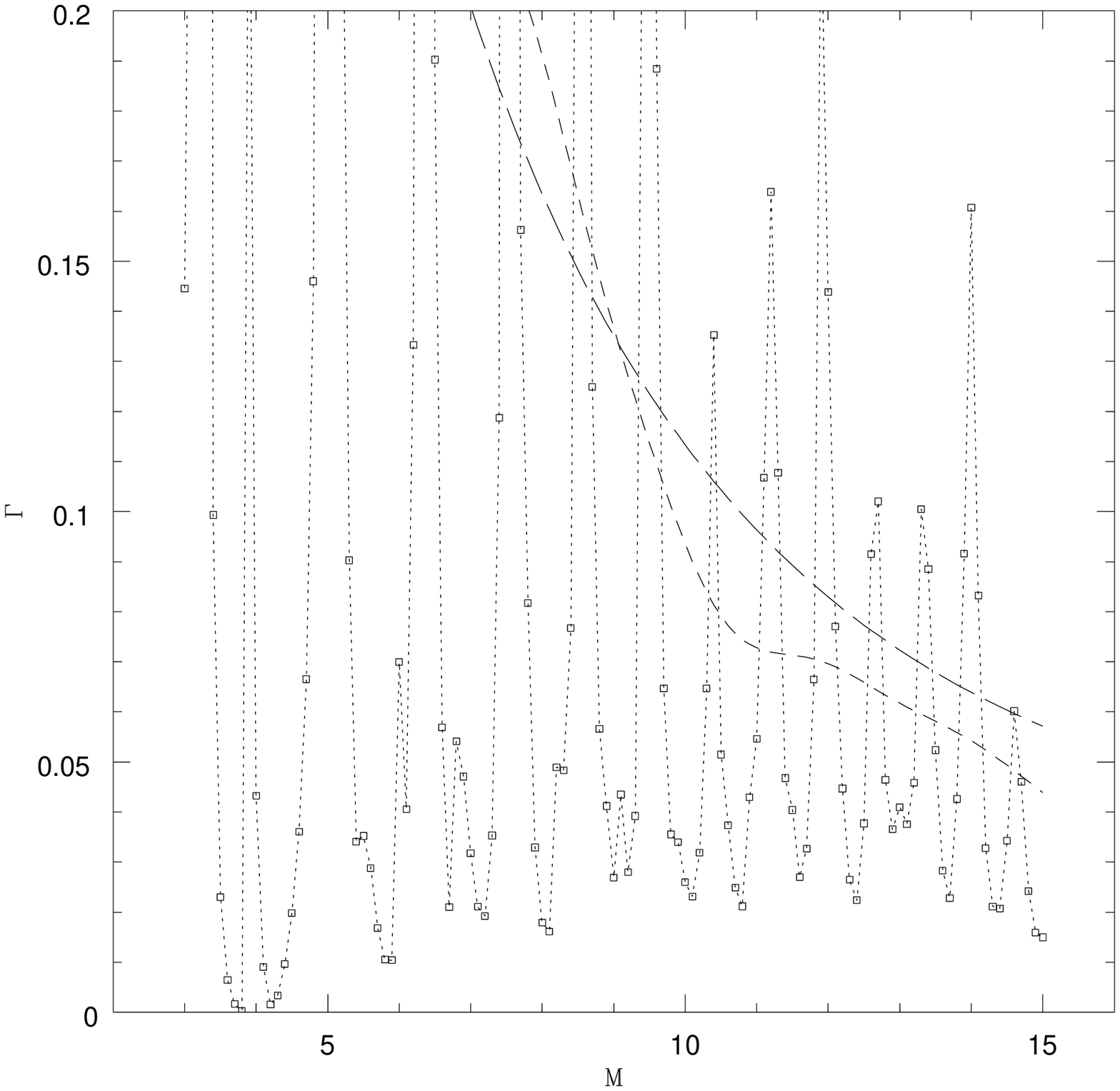}{12}{As in Fig.~10, but with
$N_c=20$ and $\Delta M=1.2$.}

The main results of this paper are exhibited in Figs.~10--13, which
we now discuss. Figures 10--12 show in dotted lines the total
annihilation width of the ``$\bar B$'' meson as a function of the
heavy quark mass $M$. Widths are presented in units of
$N_c^1 G^2 (c_V^2 - c_A^2)^2$.
The computation was carried out at the points
shown as small squares, and the dotted lines connect between them, to
guide the eye. Widths for the internal resonances have been
included, and so the plots depend on $N_c$, with $N_c=1$, 10 and 20 for
Figs.~10, 11, and 12 respectively. Also, in all three figures we show
in short dashes the Gaussian-smeared width, taking $\Delta M=0.4$, 1.2
and 1.2 for Figs.~10, 11, and 12 respectively. We see that the larger the
width of the internal resonances (that is, the smaller $N_c$), the
smoother the behavior of $\Gamma_{\rm had}$, so that $\Delta M$ as
small as 0.4 is sufficient to smooth out the $N_c=1$ case, whereas for
$N_c=20$ the larger value $\Delta M=1.2$ had to be used. In each
figure the partonic width is displayed in long dashes.

\INSERTFIG{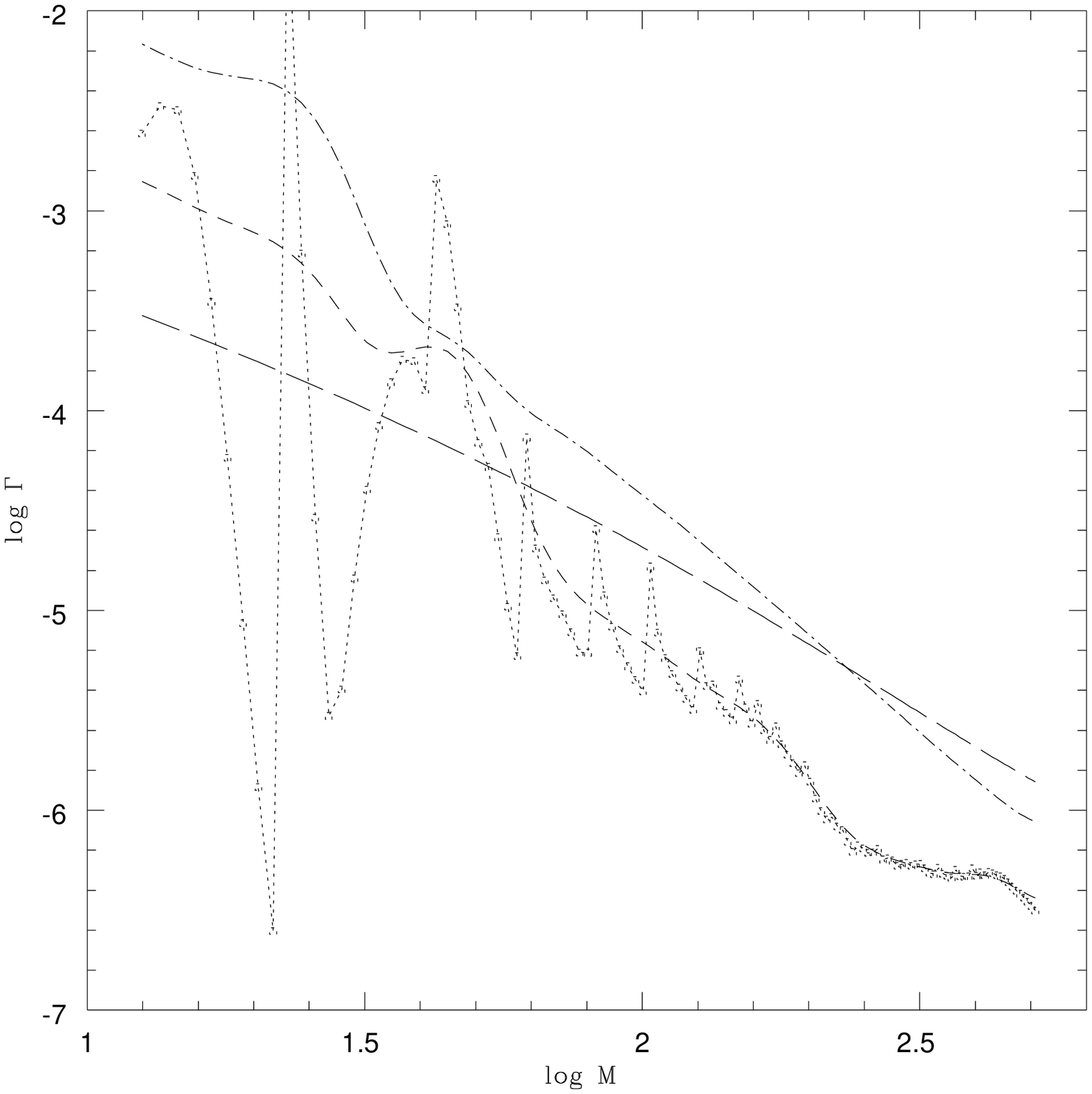}{13}{Log-log plot of the
total annihilation width
of the ``$\bar B$'' meson as a function of the heavy quark mass
$M$, for  $N_c=1$ (dotted line). The smeared
width defined in Eq.~(\ref{eq:smearing-discrete}) is shown
in short dashes, and was computed using $\Delta M=0.4$. The
partonic width is shown in long dashes. Also shown, in dash-dot-dash, is the
Gaussian-smeared result that uses the model of Eq.~(\ref{sfit}) for
the strong widths of the internal resonances.}

It is apparent that there is a large discrepancy between the actual
widths and the partonic ones. The disagreement remains after
smearing. We may now ask whether the disagreement is a correction that
decreases as a particular power of $M$. Figure 13 displays the same
information as
Fig.~10, but in log-log format. Also shown, in dash-dot-dash, is the
Gaussian-smeared result that uses the model of Eq.~(\ref{sfit}) for
the strong widths of the internal resonances.  Even though
$\Gamma_{\rm part}$ and $\Gamma_{\, \rm had}^{\, \rm avg}$ achieve
somewhat better relative agreement as $N_c$ is increased, we see that
$\Gamma_{\, \rm had}^{\, \rm avg}$
is not without structure, indicating that for this type of process
the onset of the asymptotic large $M$ limit is very delayed. This is similar to
what occurs for the decay constant $f_B$; the combination
$\sqrt{M}f_B$ is known to have large $1/M$ and $1/M^2$
corrections\cite{GM1}. We have  refrained from displaying a plot of
$L=\log(\Gamma^{{\, \rm avg}}_{{\, \rm had}}/\Gamma_{{\rm part}}-1)$, which
at large $M$ would display the leading power correction $M^{-p}$ as
the slope of $L$ versus $\log M$. The problem in doing so is that, as
discussed, it is clear that even at these large values of $M$ the
asymptotic behavior is not in sight. It is interesting to note that
the dash-dot-dash line is smooth and its slope disagrees sharply with
that of $\Gamma_{\rm part}$.

	Finally, regarding the question of a direct comparison of T
versus A widths, we remind the reader that different $N_c$ behavior
between the two widths means that, for $N_c$ sufficiently large, the A
diagram dominates.  However, there are two additional effects that
must be taken into account.  The first is that asymptotically,
$\Gamma_T \propto M$ but $\Gamma_A \propto 1/M^2$; thus for large $M$
and $N_c$ fixed, one expects $\Gamma_A/\Gamma_T \ll 1$.  Much more
interesting are the true dynamical effects obtained from the exactly
computed matrix elements and Breit-Wigner resonances in the A diagram.
Once the result for $\Gamma_T$ given in Fig.~4 of \cite{GL} is
properly normalized\footnote{In the notation of this paper, the
overall coefficient of the older figure is $N_c G^2 (c_V^2-c_A^2)^2
/\pi$.  One must divide $\Gamma_T$ by this extra $\pi$ to obtain
numerical comparisons.  There are also different CKM elements, $V_{31}
V^*_{35}$ for T and $V_{21} V^*_{25}$ for A, that we take equal for
comparison purposes.}, one still finds that $\Gamma_T$ is much larger
than $\Gamma_A$, even for fairly large $N_c$ and fairly small $M$.
For example, for $N_c = 10$ and $M=5$ (see Fig.~11) one finds
$\Gamma_T/\Gamma_A > 3$.  It appears that, in the context of the
't~Hooft model generalized to finite $N_c$, one must choose $N_c$
exceptionally large before finding values of $M$ for which $\Gamma_T
\approx \Gamma_A$.

\section{Conclusions} \label{conc}
We have studied the annihilation decays of heavy ``$\bar B$'' mesons
in the 't~Hooft model as a function of the heavy quark mass $M$ for
fixed light quark mass $m$. In the strict large $N_c$ limit the hadronic
width
solely due to annihilation decays, $\Gamma_{\rm had} = \Gamma(\bar B)$,
displays resonant
structure from intermediate light mesons. For almost all values of $M$ the
partonic width, $\Gamma_{{\rm part}}$, which displays no structure, is
at strong variance with the exact width  $\Gamma(\bar B)$.  A
comparison with a smeared version of  $\Gamma(\bar B)$ is not possible
since the resonant singularities in  $\Gamma(\bar B)$ are
non-integrable.

It seems plausible, however, that for large but finite $N_c$ and at
large enough $M$ a smeared version of  $\Gamma(\bar B)$ may agree with
$\Gamma_{{\rm part}}$. For finite $N_c$ the hadronic widths of the
resonances in $\Gamma(\bar B)$ must be included and make it possible
to study the smeared version of  $\Gamma(\bar B)$, $\Gamma^{{\, \rm
avg}}_{{\, \rm had}}$. Moreover, the leading dependences on $N_c$ of
$\Gamma^{{\, \rm avg}}_{{\, \rm had}}$  and  $\Gamma_{{\rm part}}$
coincide. However, our numerical study shows that even at masses as
large as $M=15$ in units of $g \sqrt{N_c/2\pi}$,
$\Gamma^{{\, \rm avg}}_{{\, \rm
had}}$  and  $\Gamma_{{\rm part}}$ disagree significantly.
Although increasing $N_c$ improves the relative agreement somewhat, the
improvement is not strong enough to claim the onset of asymptotic scaling,
even by $M=15$.

Our analysis does not conclusively show the breakdown of duality for
annihilation decays of heavy mesons. But it seems clear that if
duality is operative asymptotically, it must be that asymptotia is much
delayed for these types of decays.

\vskip1.2cm
{\it Acknowledgments}
\hfil\break
RFL is pleased to acknowledge enlightening discussions with Nathan
Isgur.  This work is supported by the Department of Energy under
contract Nos.\ DOE-FG03-97ER40506 and DE-AC05-84ER40150.

\appendix
\section{van Royen-Weisskopf Relation in Arbitrary Dimensions}
	It is interesting to consider the generalization of
the van Royen-Weisskopf relation\cite{vRW}, which connects the meson decay
constant to the value of the meson wavefunction at zero quark
separation, and was therefore implicitly used in the original naive
argument in Sec.~\ref{intro} that the annihilation diagram is
suppressed compared to the tree diagram.  This relation is proved
using nonrelativistic constituent quarks within the meson;
nevertheless, one may consider its generalization to arbitrary $N_c$
and $D$ spacetime dimensions, in which it reads
\begin{equation}
f_B^2 = \frac{4N_c |\psi(0)|^2}{M+m} .
\end{equation}
Note particularly that $D$ only enters this expression implicitly in
$\psi(0)$.  The explicit factor of $N_c$ from $f_B^2$ is in fact the
source of the enhancement of the A width to the T width.  To proceed,
we require a model for the wavefunction, for which we choose
\begin{equation}
\psi({\bf r}) = R(r) Y_{00} (\Omega) = N e^{-\mu r} Y_{00},
\end{equation}
where $\mu$ is a typical hadronic mass scale, the $D$-dimensional
spherical harmonic is given by
\begin{equation}
|Y_{00} (\Omega)|^2 = \frac{1}{\int d \Omega} = \frac{\Gamma
((D-1)/2)}{2\pi^{(D-1)/2}},
\end{equation}
and $N^2 = (2\mu)^{D-1}/\Gamma(D-1)$.  We then have
\begin{equation}
f_B^2 = \frac{4N_c}{M+m} \cdot \frac{\Gamma((D-1)/2)}{2\pi^{(D-1)/2}}
\cdot \frac{(2\mu)^{D-1}}{\Gamma(D-1)} .
\end{equation}
Note that the mass dimension of $f_B$ is $M^{D/2-1}$, as can also be
shown directly from its definition (\ref{eq:decayconstant}) as a
matrix element.

	In the context of dimensional regularization, the factor
appearing with the inclusion of each additional loop is
$(4\pi)^{-D/2}$.  We must also, according to the arguments of
Sec.~\ref{intro}, divide out powers of $m_B$ ($= M+m$ in this model)
to obtain a dimensionless ratio.  The relevant ratio between the A and
T diagram widths in this simple model is thus given by
\begin{equation}
(4\pi)^{D/2} \frac{f_B^2}{m_B^{D-2}} = N_c \left( \frac{\mu}{M+m}
\right)^{D-1} \cdot 4^D \sqrt{\pi} \, \frac{\Gamma((D-1)/2)}{\Gamma(D-1)}
.
\end{equation}
Removing the explicit $N_c$ and the mass ratio, the remaining
$D$-dependent coefficient actually reaches a maximum for $D=9$.  We
see that even in this simple model --- no dynamics, not even a
helicity suppression factor, has been included --- tells us the ratio
between the A and T diagrams depends sensitively on interplay between
the value of $N_c$, the quark masses and interaction energies, and the
number of spacetime dimensions.

\newpage
\begin{center}
{\bf Figure Captions}
\end{center}

FIG.\ 1. The (color-unsuppressed) ``tree'' (T) parton diagram for the
decay of one meson into two mesons.  Ovals indicate the binding of
partons into hadrons.

FIG.\ 2. The ``color-suppressed'' (C) parton diagram for the decay of
one meson into two mesons.  Ovals indicate the binding of partons into
hadrons.

FIG.\ 3. The ``annihilation'' (A) parton diagram for the decay of one
meson into two mesons.  Ovals indicate the binding of partons into
hadrons.

FIG.\ 4$a$. Diagram giving rise to the ``tree'' amplitude of Fig.~1
upon a vertical cut through the center (application of unitarity).
The vertex blobs indicate $W$ exchange.

FIG.\ 4$b$. Diagrams giving rise to the ``annihilation'' amplitude of
Fig.~3 upon a vertical cut through the center (application of
unitarity).  The vertex blobs indicate $W$ exchange.  Strongly
produced $q\bar q$ pairs are not drawn here for simplicity.

FIG.\ 5$a$. Electroweak ``exchange'' (E) parton diagram.

FIG.\ 5$b$. ``Penguin'' (P) parton diagram.

FIG.\ 5$c$. ``Penguin annihilation'' (PA) parton diagram.  Since the
initial and final states are color singlets, the intermediate state
actually requires at least two gluons; however, the archetype
presented here exhibits the same $N_c$ counting.

FIG.\ 6$a$. Diagram for ``tree'' (T) meson exclusive decay.  Numbers
indicate quark labels used in the text (except {\bf 0}, which refers to the
ground-state ``$\bar B$'' meson), while letters indicate the
eigenvalue index of meson resonances.  One can also consider
contact-type diagrams, in which the point labeled by\/ {\bf n} is not
coupled to a resonance.

FIG.\ 6$b$. Diagram for ``annihilation'' (A) meson exclusive decay.
Numbers indicate quark labels used in the text (except {\bf 0}, which
refers to the ground-state ``$\bar B$'' meson), while letters indicate
the eigenvalue index of meson resonances.  One can also consider
contact-type diagrams, in which the point labeled by\/ {\bf p} is not
coupled to a resonance.

FIG.\ 7.  Feynman diagram topology for the interference between A and
T amplitudes.

FIG.\ 8.  Ratio of the total width, solely
from the annihilation topology, computed for $K=50$ to that for
$K=200$, in the case of ($a$) $N_c=10$, and ($b$) $N_c=1$. In both
cases the ratio of the Gaussian-smeared widths is also shown, with a
Gaussian width of  $\Delta M=1.2$ in ($a$) and  $\Delta M=0.4$ in ($b$).
The unit of mass here and in all subsequent figures is $g\sqrt{N_c/2\pi}$.

FIG.\ 9.  Meson strong decay widths in units of
$g \sqrt{N_c/2\pi} \cdot 1/N_c$ as computed via (\ref{swidth}).
The smooth dashed curve is the fit function (\ref{sfit}).

FIG.\ 10. Total annihilation width
of the ``$\bar B$'' meson as a function of the heavy quark mass
$M$, for  $N_c=1$ (dotted line). The smeared
width defined in Eq.~(\ref{eq:smearing-discrete}) is shown
in short dashes, and was computed using $\Delta M=0.4$. The
partonic width is shown in long dashes.
Units of the widths here and below are $N_c^1 G^2 (c_V^2 - c_A^2)^2$.

FIG.\ 11. As in Fig.~10, but with
$N_c=10$ and $\Delta M=1.2$.

FIG.\ 12. As in Fig.~10, but with
$N_c=20$ and $\Delta M=1.2$.

FIG.\ 13. Log-log plot of the
total annihilation width
of the ``$\bar B$'' meson as a function of the heavy quark mass
$M$, for  $N_c=1$ (dotted line). The smeared
width defined in Eq.~(\ref{eq:smearing-discrete}) is shown
in short dashes, and was computed using $\Delta M=0.4$. The
partonic width is shown in long dashes. Also shown, in dash-dot-dash, is the
Gaussian-smeared result that uses the model of Eq.~(\ref{sfit}) for
the strong widths of the internal resonances.


\begin{thebibliography}{199}

\bibitem{GL}B. Grinstein and R. F. Lebed, Phys.\ Rev.\ D {\bf 57},
1366 (1998).

\bibitem{Wils}K. Wilson, Phys.\ Rev.\ {\bf 179}, 1499 (1969) and
Phys.\ Rev.\ D {\bf 3}, 1818 (1971);\\
W. Zimmerman, Ann.\ Phys.\ (New York) {\bf 77}, 536 and 570, 1973.

\bibitem{DIS}H. Georgi and H. D. Politzer, Phys.\ Rev.\ D {\bf 9}, 416
(1974);\\
D. J. Gross and F. Wilczek, Phys.\ Rev.\ D {\bf 9}, 920 (1974).

\bibitem{SVZ}M. A. Shifman, A. I. Vainshtein, and V. I. Zakharov,
Nucl.\ Phys.\ B {\bf 147}, 385 (1979); {\it ibid.\/}, 448.

\bibitem{CGG}J. Chay, H. Georgi, and B. Grinstein, Phys.\ Lett.\ B {\bf
247}, 399 (1990).

\bibitem{bigi}I. I. Bigi, B. Blok, M. A. Shifman, N. G. Uraltsev, and A.
Vainshtein, in Proceedings of {\it The Fermilab Meeting: DPF 92},
Edited by Carl H. Albright, Peter H. Kasper, Rajendran Raja, and John
Yoh. (World Scientific, River Edge, N.J., 1993), p. 610.

\bibitem{tLam}ALEPH Collaboration (R. Barate {\it et al.}), Eur.\
Phys.\ C {\bf 2}, 197 (1998);\\
ALEPH Collaboration (R. Barate {\it et al.}), Preprint No.\ CERN
PPE/97 157 (unpublished);\\
Particle Data Group (R. M. Barnett {\it et al.}), Phys.\ Rev.\ D {\bf
54}, 1 (1996).

\bibitem{NS}M. Neubert and C. T. Sachrajda, Nucl.\ Phys.\ B {\bf 483},
339 (1997).

\bibitem{tH}G. 't~Hooft, Nucl.\ Phys.\ B {\bf 75}, 461 (1974).

\bibitem{CCG}C. G. Callan, Jr., N. Coote, and D. J. Gross, Phys.\
Rev.\ D {\bf 13}, 1649 (1976).

\bibitem{tHN}G. 't~Hooft, Nucl.\ Phys.\ B {\bf 72}, 461 (1974).

\bibitem{Wit}E. Witten, Nucl.\ Phys.\ B {\bf 160}, 57 (1979) provides
an excellent review of these properties.

\bibitem{GHLR}M. Gronau, O. F. Hern\'{a}ndez, D. London, and J. L.
Rosner, Phys.\ Rev.\ D {\bf 50}, 4529 (1994).

\bibitem{JM}R. L. Jaffe and P. F. Mende, Nucl.\ Phys.\ B {\bf 369},
189 (1992).

\bibitem{GM1}B. Grinstein and P. F. Mende, Phys.\ Rev.\ Lett.\ {\bf
69}, 1018 (1992).

\bibitem{GM2}B. Grinstein and P. F. Mende, Nucl.\ Phys.\ B {\bf 425},
451 (1994).

\bibitem{Ein}M. B. Einhorn, Phys.\ Rev.\ D {\bf 14}, 3451 (1976).

\bibitem{Mul}A. J. Hanson, R. D. Peccei, and M. K. Prasad, Nucl.\
Phys.\ B {\bf 121}, 477 (1977); K. Karamchetti, {\it Principles of
Ideal-Fluid Aerodynamics}, John Wiley \& Sons, New York, 1966.

\bibitem{Blok}B. Blok, M. Shifman, and D.-X. Zhang, Phys.\ Rev.\ D
{\bf 57}, 2691 (1998);\\
B. Blok, Nucl.\ Phys.\ Proc.\ Suppl.\ {\bf 64}, 481 (1998).  

\bibitem{Brow}R. C. Brower, W. L. Spence, and J. H. Weis, Phys.\ Rev.\
D {\bf 19}, 3024 (1979).

\bibitem{vRW}R. van Royen and V. F. Weisskopf, Nuovo Cimento {\bf 50A},
617 (1967).

\end{thebibliography}
\end{document}